\pdfoutput=1
\documentclass[prd,onecolumn,notitlepage,eqsecnum,nofootinbib,floatfix]{revtex4-1}
\usepackage{amssymb}
\usepackage{amsmath}
\usepackage{bm}
\usepackage{graphicx}

\begin{document}
\title{Self-force on a charge outside a five-dimensional black hole}   
\author{Matthew J.\ S.\ Beach, Eric Poisson, and Bernhard G.\ Nickel} 
\affiliation{Department of Physics, University of Guelph, Guelph,
  Ontario, N1G 2W1, Canada} 
\date{May 14, 2014} 
\begin{abstract} 
We compute the electromagnetic self-force acting on a charged particle  
held in place at a fixed position $r$ outside a five-dimensional black 
hole described by the Schwarzschild-Tangherlini metric. Using a
spherical-harmonic decomposition of the electrostatic potential and a
regularization prescription based on the Hadamard Green's function, we
express the self-force as a convergent mode sum. The self-force is 
first evaluated numerically, and next presented as an analytical
expansion in powers of $R/r$, with $R$ denoting the event-horizon
radius. The power series is then summed to yield a closed-form
expression. Unlike its four-dimensional version, the self-force
features a dependence on a regularization parameter $s$ that
can be interpreted as the particle's radius. The self-force is
repulsive at large distances, and its behavior is related to a model
according to which the force results from a gravitational interaction
between the black hole and the distribution of electrostatic field
energy attached to the particle. The model, however, is shown to
become inadequate as $r$ becomes comparable to $R$, where the
self-force changes sign and becomes attractive. We also calculate the
self-force acting on a particle with a scalar charge, which we find to
be everywhere attractive. This is to be contrasted with its
four-dimensional counterpart, which vanishes at any $r$.     
\end{abstract}  
\maketitle

\section{Introduction and summary} 
\label{sec:intro} 

A particle held in place at a fixed position $r$ outside a nonrotating
black hole of mass $M$ requires an external agent to supply an
external force $F_{\rm ext}$ that compensates for the black hole's
gravity. When the particle carries an electric charge $q$, the
external force is smaller than when the particle is neutral. The 
difference is accounted for by the particle's electromagnetic
self-force, which originates in a subtle interaction between the
particle, the electric field it generates, and the spacetime curvature
around the black hole. The electromagnetic self-force acting on a
charged particle at rest outside a Schwarzschild black hole was
computed by Smith and Will \cite{smith-will:80}.
The only nonvanishing component of the force vector $F^\alpha$ is the
radial component $F^r$, and the force invariant 
$F := \pm\sqrt{ g_{\alpha\beta} F^\alpha F^\beta }$, with the sign 
adjusted so that the sign of $F$ agrees with the sign of $F^r$, is
given by 
\begin{equation} 
F = \frac{q^2 R}{2 r^3}, 
\label{eq:Smith-Will} 
\end{equation} 
where $R = 2M$ is the event-horizon radius (we use geometrized units,
in which $G=c=1$). The positive sign on the right-hand side indicates
that the self-force is repulsive, which leads to a smaller 
$F_{\rm ext}$ when the particle carries a charge.  

The repulsive nature of the electromagnetic self-force is a surprising
feature that is difficult to explain. An attempt to provide some
intuition relies on the fact that the event horizon is necessarily an
equipotential surface, which suggests that the black hole should
behave as a perfect conductor. This observation leads to the
expectation that the self-force could be derived (up to numerical
factors) on the basis of an elementary model involving a spherical
conductor of radius $R$ in flat spacetime. The model features a charge
$q$ at position $r$, a first image charge $q' = -q(R/r)$ at position
$R^2/r$ inside the conductor, and a second image charge $-q' = q(R/r)$
at the center. The first image charge produces a grounded conductor
with a net charge $q'$ distributed on its surface, and the second
image charge eliminates this net charge, without violating the
equipotential condition at the surface. (The black hole does not
support a net charge.) The model predicts a self-force resulting from
an interaction between the charge $q$ and the image dipole inside the
conductor, and a simple calculation neglecting corrections of order
$(R/r)^2$ reveals that the self-force scales as $F \sim -q^2 R^3/r^5$,
with a negative sign indicating an attractive force. The model is a
complete failure: it fails to produce to the correct scaling with $R$
and $r$, and it even fails to produce the correct sign.

Another attempt to provide intuition, proposed in Sec.~IV of
Ref.~\cite{burko-etal:00}, produces a more intelligible
picture. This model focuses its attention on the force acting on the 
black hole instead of the force acting on the charged particle. This
force is necessarily gravitational in nature, and according to
Newton's third law, it must be equal in magnitude to the force acting
on the particle. (The model features a mixture of Newtonian and
relativistic ideas.) The force on the black hole is produced in part
by the particle's mass, but there is also a contribution from the 
distribution of electrostatic field energy that surrounds the
charge. In this view, the charged particle behaves as an infinitely 
extended body, and the black hole is comparatively much smaller.  
The force on the black hole is then $F_{\rm hole} = -M m(r)/r^2$, with
the negative sign indicating that the force is attractive, and $m(r)$
denoting the mass within radius $r$ associated with the particle and
the distribution of field energy. The particle's mass $m$ is
identified with $m(\infty)$, and 
$m(r) = m - \int_r^\infty 4\pi \rho r^2\, dr$, where  
$\rho = E^2/(8\pi)$ is the density of field energy. With $E = q/r^2$
we have that $m(r) = m - q^2/(2r)$, and the force becomes   
$F_{\rm hole} = -M m/r^2 + q^2 R/(4 r^3)$. The first term is the
attractive gravitational force exerted by the particle, and the second
term is a repulsive contribution from the field energy. Writing 
$F_{\rm charge} = F_{\rm hole}$ gives us an alternative
interpretation: the first term is the gravitational force 
exerted by the black hole, and the second term is the self-force. This
model is a success: it produces the correct scaling with $R$ and $r$,
and it produces the correct sign. It reproduces the Smith-Will force
of Eq.~(\ref{eq:Smith-Will}) up to a factor of 2.  

The failure of the electrostatic model at providing a reliable
expression for the self-force has been a source of fascination in the
literature, and it has motivated a line of inquiry that probes into
the mysteries of the self-force in various circumstances. Thus,
authors have replaced the black hole with various material bodies
\cite{unruh:76,burko-etal:00, shankar-whiting:07, 
  drivas-gralla:11, isoyama-poisson:12}; they observed that the
Smith-Will behavior of Eq.~(\ref{eq:Smith-Will}) is universal at large
distances, but modified when $r$ becomes comparable to the body's
radius $R$. Other authors have replaced the asymptotically-flat
boundary conditions of the Schwarzschild spacetime by asymptotic 
cosmological conditions (specifically, de Sitter or anti de Sitter
conditions \cite{kuchar-poisson-vega:13}); they observed that the
Smith-Will behavior continues to hold approximately when the
black-hole and cosmological scales are well separated, but is
substantially modified when the scales are comparable.  

In this paper we continue this line of inquiry, and ask whether the
interpretation of the self-force as a gravitational interaction
between the black hole and the electrostatic field energy attached to
the particle continues to apply in higher dimensions. Extending the
model to an $(n+2)$-dimensional spacetime, with $n$ denoting the
number of angular directions, we have that the force on the black hole
is now given by $F_{\rm hole} = -Mm(r)/r^n$. The mass within radius
$r$ becomes $m(r) = m - \int_r^\infty \Omega_n \rho r^n\, dr$, where  
$\rho = E^2/(2\Omega_n)$ is the density of field energy, and
$\Omega_n$ is the area of a unit $n$-sphere. With $E = q/r^n$ we have
that $m(r) = m - q^2/[2(n-1) r^{n-1}]$, and we obtain 
$F_{\rm hole} = -M m/r^n + q^2 M/[2(n-1) r^{2n-1}]$. The second term
is identified with the electromagnetic self-force, and relating the
black-hole mass to its event-horizon radius $R$ via 
$M = \frac{1}{2}(n-1) R^{n-1}$, we arrive at an expected scaling of  
$q^2 R^{n-1}/(4r^{2n-1})$ for the self-force. We wish to know whether 
this expectation is borne out by an actual computation. Self-forces in
higher-dimensional spacetimes were also considered by Frolov and
Zelnikov \cite{frolov-zelnikov:12a, frolov-zelnikov:12b,
  frolov-zelnikov:12c, frolov-zelnikov:12d}, who provided concrete
results for the specific case of Majumdar-Papapetrou spacetimes.    

\begin{figure} 
\includegraphics[width=0.8\linewidth]{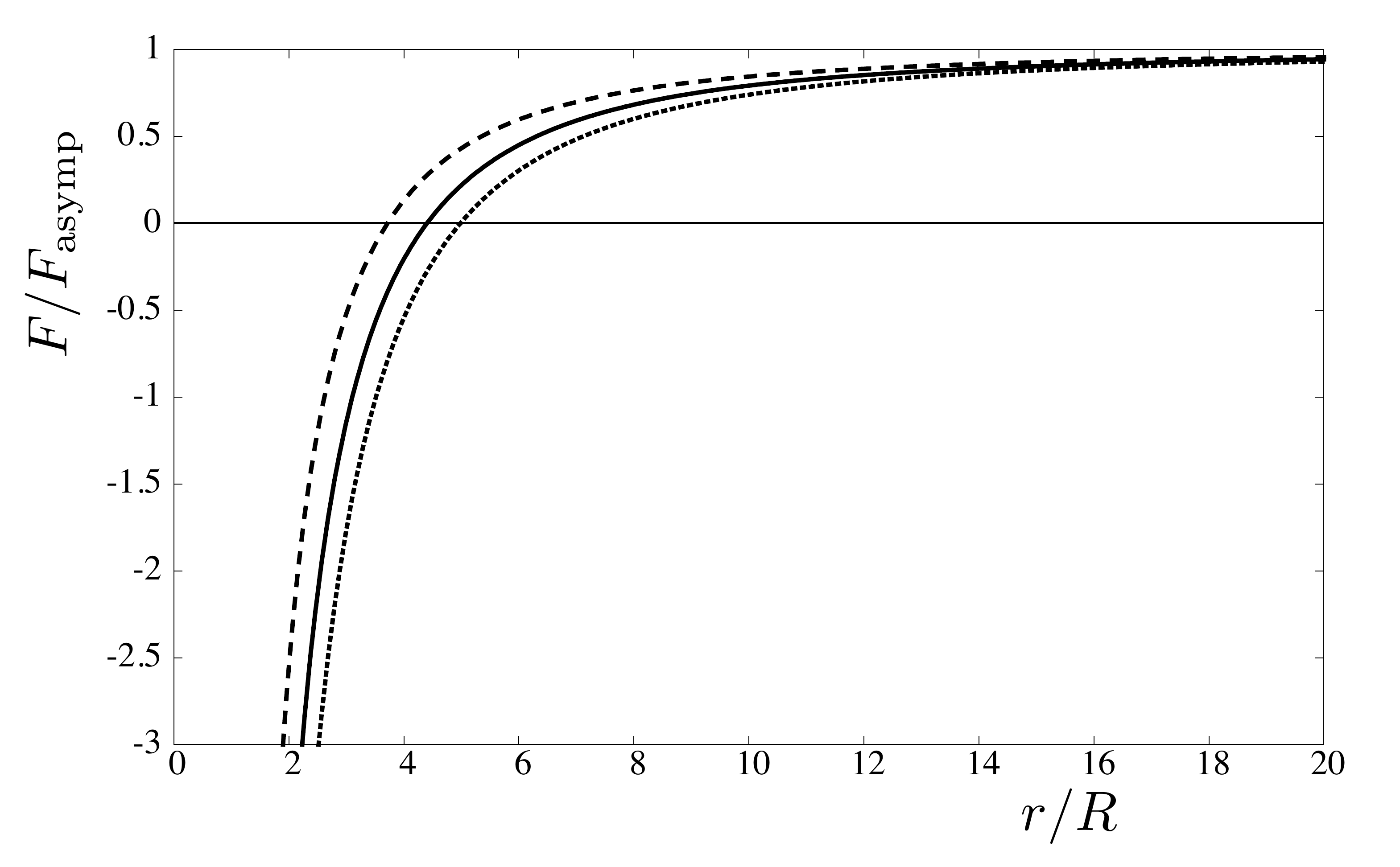}
\caption{Electromagnetic self-force acting on a particle of charge $q$
  at a fixed position $r$ in the five-dimensional
  Schwarzschild-Tangherlini spacetime. The self-force $F$ is divided
  by its asymptotic expression $F_{\rm asymp} = q^2 R^2/(2 r^5)$, and
  it is plotted as a function of $r/R$ for $\tilde{s} = 10^{-4}$
  (long-dashed curve), $\tilde{s} = 10^{-6}$ (solid curve), and
  $\tilde{s} = 10^{-8}$ (short-dashed curve). The self-force is
  positive (repulsive) when $r/R$ is sufficiently large, but it
  changes sign and becomes attractive when $r/R$ becomes comparable to
  4. For $\tilde{s} = 10^{-4}$ the transition occurs at $r/R \simeq
  3.7154$, for $\tilde{s} = 10^{-6}$ it occurs at $r/R \simeq 4.4010$,
  and for $\tilde{s} = 10^{-8}$ it occurs at $r/R \simeq 4.9830$.}    
\label{fig:emsf1} 
\end{figure}

For reasons that will be explained below, our calculation of the
self-force is restricted to the five-dimensional case. We obtain 
\begin{equation} 
F = \frac{q^2 R^2}{2 r^5}\, \frac{\Xi}{f^{3/2}}, 
\label{eq:F_5D} 
\end{equation} 
where $R$ is the event-horizon radius, $f := 1-(R/r)^2$, and 
\begin{equation} 
\Xi = - \frac{1}{4x} + \frac{5}{8} + \frac{139}{96}\, x 
- \frac{281}{192}\, x^2 + \biggl( \frac{1}{4x} + \frac{1}{2} 
- \frac{15}{16}\, x \biggr) \sqrt{f} + \frac{3}{16}\, x (6-5x) 
  \ln \frac {\tilde{s} x (1+\sqrt{f})}{8 \sqrt{f}} 
\end{equation}
with $x := (R/r)^2$. The self-force depends on an unknown
parameter, the dimensionless quantity $\tilde{s} := s/R$, which
originates in the regularization prescription to be described
below. An interpretation for the length scale $s$ is that it
represents the radius of the particle, which must of course be much
smaller than the black hole, so that $\tilde{s} \ll 1$. The
self-force, therefore, is not independent of the particle's size, and
presumably this is an indication that in five dimensions, the
self-force cannot be expected to be independent of the details of
internal structure.  A graph of $\Xi/f^{3/2}$ for selected values of 
$\tilde{s}$ is displayed in Fig.~\ref{fig:emsf1}.    

When $r$ is large compared with $R$, the function $\Xi/f^{3/2}$
behaves as $1 + O(x)$, and the self-force becomes 
\begin{equation} 
F \sim \frac{q^2 R^2}{2 r^5}, \qquad (r \gg R).   
\end{equation} 
This repulsive behavior matches the expectation from the gravitational
model, up to a factor of two that was also seen in the
four-dimensional case. When $r$ decreases toward $R$, however, the
self-force force changes sign and becomes attractive. As $r$
approaches $R$ the diverging factor $f^{-3/2}$ begins to dominate, but
the divergence is limited by the fact that the particle cannot be
closer to the horizon than a distance of order $s$. Taking 
$r/R > 1 + \tilde{s}$, we find that the self-force is bounded by  
\begin{equation} 
F > -\frac{3}{128\sqrt{2}} \frac{q^2}{R^3} 
\frac{1}{\tilde{s}^{3/2}} \ln \frac{128}{\tilde{s}}. 
\label{eq:F_horizonbound} 
\end{equation} 
In spite of this bound, the behavior of the self-force very close to
the horizon should be viewed with suspicion, because a large $|F|$
implies a large electric field that can no longer be treated as
a test field in a fixed background spacetime. The detailed description
of the self-force reveals that the interpretation in terms of a
gravitational interaction does not hold up to five-dimensional
scrutiny. While the large-$r$ behavior of the self-force is repulsive
and compatible with the model, the agreement does not persist when $r$
becomes comparable to $R$. 

Unlike Smith and Will, our computation of the five-dimensional
self-force does not proceed on the basis of an exact solution to
Maxwell's equations for a point charge in the
Schwarzschild-Tangherlini spacetime. Indeed, such a five-dimensional
analogue of the Copson solution \cite{copson:28, linet:76} is not
known. Our method of calculation is therefore more convoluted. We 
begin in Sec.~\ref{sec:EM} with the formulation of Maxwell's equations
in higher-dimensional spacetimes, their specialization to the specific
case of a point charge in the Schwarzschild-Tangherlini spacetime, and
a presentation of the solution in terms of a decomposition in
higher-dimensional spherical harmonics. This leads to a self-force 
expressed as an infinite and diverging sum over spherical-harmonic
modes.  

The mode-sum evaluation of the self-force requires regularization, and
we carry out the necessary steps in Sec.~\ref{sec:Hadamard}. We adopt
a regularization prescription based on Hadamard's Green's function
\cite{hadamard:23}, a local expansion of the electrostatic potential
that identifies the singular part that must be subtracted before the
mode sum is evaluated. While Hadamard regularization can be
formulated in any spacetime dimension, its practical implementation
becomes increasingly difficult as the number of dimensions increases,
for the simple reason that the electric field of a point charge
becomes increasingly singular at the position of the particle. With
the techniques at our disposal we were able to handle the
five-dimensional case with relative ease, and this motivated our
restriction to five dimensions. An extension to higher dimensions is
possible, but would require a substantial amount of additional work. 
 
Unlike the situation in four dimensions, the five-dimensional Hadamard
Green's function features a logarithmic dependence on the separation
between field and source points. This is the source of the $\ln(s/R)$
term in the self-force, with the length parameter $s$ interpreted as
the particle's radius. Unlike its four-dimensional version, the
five-dimensional self-force depends on the details of the particle's
internal structure. This dependence is likely to be even more dramatic 
in higher dimensions, because the field of a point charge becomes
increasingly singular and requires additional regularization. It would
be interesting to pursue these matters by performing a self-force
calculation in six dimensions.    

The calculation of the self-force proceeds in Sec.~\ref{sec:SF} with a
numerical evaluation of the regularized mode sum, and an
analytical evaluation presented as an expansion in powers of $R/r$. We
carry out this expansion to a very high order, and manage to sum the
series to the closed-form expression displayed in Eq.~(\ref{eq:F_5D}).   

In Sec.~\ref{sec:scalar} we exploit the same methods to calculate the
self-force acting on a scalar charge $q$ at a fixed position $r$ in
the five-dimensional Schwarzschild-Tangherlini spacetime.  Our final
result is displayed in Eq.~(\ref{eq:scalar_sf}) below. When $r$ is
much larger than $R$ we find that the scalar self-force behaves as      
\begin{equation} 
F_{\rm scalar} \sim -\frac{3}{8} \frac{q^2 R^4}{r^7}  
\ln \frac{2r}{\sqrt{\tilde{s}} R}. 
\end{equation} 
The self-force is attractive everywhere, and its scaling with 
$R^4 \ln r /r^7$ can be contrasted with the $R^2/r^5$ scaling of the 
electromagnetic self-force. This result can also be contrasted with
Wiseman's four-dimensional expression \cite{wiseman:00}: 
$F_{\rm scalar} = 0$. Like its electromagnetic counterpart, the scalar
self-force is bounded by Eq.~(\ref{eq:F_horizonbound}) when the
particle is close to the horizon.  

What would happen to the five-dimensional self-force if the
topology of the event horizon were changed from the $S^3$ topology
examined here to the $S^2 \times \mathbb{R}$ topology of a black
string? The regularization techniques developed in this paper could be
adapted to this new situation, and a fresh calculation of the
self-force could be attempted. Would the self-force continue to  
diverge as the event horizon is approached? We hope to return to this
question in future work. In the remainder of the paper we present the
detailed calculations that lead to the results summarized in this
introductory section.    

\section{Electrostatics in a higher-dimensional black-hole spacetime}  
\label{sec:EM} 

In this section we formulate Maxwell's equations in a curved spacetime
of arbitrary dimensionality, and specialize them to the description of a
static electric charge in the spacetime of a nonrotating black
hole. This spacetime is static and spherically symmetric, and we
denote the number of angular directions by $n$; the total number of
spatial dimensions is then $n+1$, and $n+2$ is the number of spacetime 
dimensions. 

\subsection{Maxwell's equations and Lorentz force} 

Maxwell's equations in a curved, $(n+2)$-dimensional spacetime are
expressed in covariant form as 
\begin{equation} 
\nabla_\beta F^{\alpha\beta} = \Omega_n j^\alpha, \qquad 
\nabla_{[\alpha} F_{\beta\gamma]} = 0,  
\label{eq:Maxwell} 
\end{equation} 
where $F^{\alpha\beta}$ is the electromagnetic field tensor,
$j^\alpha$ is the current density, $\nabla_\alpha$ is the covariant
derivative operator, and $\Omega_n$ is the area of an
$n$-dimensional unit sphere --- an explicit expression is given in
Eq.~(\ref{eq:volume}); indices enclosed within square brackets are
antisymmetrized. The sourcefree Maxwell equations can be solved by
expressing the electromagnetic field in terms of a vector potential,
\begin{equation} 
F_{\alpha\beta} = \nabla_\alpha \Phi_\beta 
- \nabla_\beta \Phi_\alpha. 
\end{equation} 
In a given Lorentz frame in flat spacetime, the components of a static
electric field are given by $E_a = F_{at} = \partial_a \Phi$ with
$\Phi := \Phi_t$, and Maxwell's equations reduce to Gauss's law 
$\bm{\nabla} \cdot \bm{E} = \Omega_n j^t$, where $j^t$ is the
charge density. The field produced by a point charge $q$ at the
spatial origin of the coordinate system is given by 
$\bm{E} = (q/r^n) \bm{\Omega}$, where $\bm{\Omega} := \bm{x}/r$ is a
unit vector in the direction of the field point $\bm{x}$, and the
associated potential is given by $(n-1)\Phi = -q/r^{n-1}$. 

The current density of a point charge $q$ moving on a world line
described by the parametric relations $z^\alpha(\tau)$ (with $\tau$
denoting proper time) is given by   
\begin{equation} 
j^\alpha(x) = q \int u^\alpha \delta\bigl(x,z(\tau)\bigr)\, d\tau, 
\label{eq:density} 
\end{equation} 
where $\delta(x,x')$ is a scalarized Dirac distribution defined by
$\int a(x') \delta(x,x') \sqrt{-g'}\, d^{n+2}x' = a(x)$ when $x$ lies
within the domain of integration; $g'$ is the metric determinant
evaluated at $x'$, and $a(x')$ is an arbitrary test function.  

Formally, the electromagnetic self-force acting on this point charge
is given by the Lorentz force 
\begin{equation} 
F^\alpha = q F^\alpha_{\ \beta} u^\beta, 
\label{eq:Lorentz} 
\end{equation} 
where $F_{\alpha\beta}$ is the electromagnetic field produced by the 
charge. Since this field diverges at the position of the particle, the
equation has only formal validity, and the field must be regularized
before the self-force is computed.  

\subsection{Schwarzschild-Tangherlini spacetime} 

We specialize the general formulation of Maxwell's equations to the
case of a charge $q$ held at a fixed position in a  
higher-dimensional analogue of the Schwarzschild spacetime, often
named the Tangherlini spacetime \cite{tangherlini:63}. Its metric is
given by  
\begin{equation} 
ds^2 = -f\, dt^2 + f^{-1}\, dr^2 + r^2\, d\Omega^2_n, 
\label{eq:STmetric}
\end{equation} 
where 
\begin{equation}
f := 1 - (R/r)^{n-1} 
\end{equation} 
and $d\Omega^2_n := \Omega_{AB}\, d\theta^A d\theta^B$ is the metric
on a unit $n$-sphere --- refer to the Appendix for a fuller
description of the notation employed here and below. The
gravitational radius $R$ marks the position of the event horizon, and
it is related to the gravitational (ADM) mass $M$ by 
$M = \frac{1}{2}(n-1) R^{n-1}$. 

The fixed position of the particle is described by $r=r_0$ and
$\theta^A = \theta^A_0$. To condense the notation it is helpful to
represent the angular coordinates $\theta^A$ by a unit vector
$\bm{\Omega}$ defined in such a way that the relation between the
spherical polar coordinates $(r,\theta^A)$ and quasi-Cartesian
coordinates $x^a$ is given by the usual 
$\bm{x} = r \bm{\Omega}(\theta^A)$. In this notation the variable
position of a point on a $t=\mbox{constant}$ hypersurface is
represented by $(r,\bm{\Omega})$, and the fixed position of the charge
is designated by $(r_0,\bm{\Omega}_0)$.    
   
For this static situation the only nonvanishing component of the
vector potential is $\Phi := \Phi_t$, and Maxwell's equations reduce
to the single equation  
\begin{equation} 
r^2 \partial_{rr} \Phi + n r \partial_r \Phi + \frac{1}{f} D^2 \Phi 
= r^2 \Omega_n j^t, 
\label{eq:Maxwell1} 
\end{equation} 
in which $\partial_r$ denotes a partial derivative with respect to
$r$, and $D^2$ is the Laplacian operator on the unit $n$-sphere (refer
to the Appendix). The charge density $j^t$ can be obtained
from the general expression of Eq.~(\ref{eq:density}) by switching
integration variables from $\tau$ to $z^0(\tau)$. We get
\begin{equation} 
j^t = q \frac{\delta(r-r_0)}{r_0^n} \delta(\bm{\Omega},\bm{\Omega}_0),
\end{equation} 
where $\delta(\bm{\Omega},\bm{\Omega}_0)$ is the angular Dirac
distribution introduced in Eq.~(\ref{eq:delta_angular}). 

A formal expression for the self-force acting on the charged particle
can be obtained from Eq.~(\ref{eq:Lorentz}). Its only nonvanishing
component is   
\begin{equation} 
F^r = q \sqrt{f_0}\, \partial_r \Phi(r_0, \bm{\Omega}_0),
\end{equation} 
where $f_0 := 1-(R/r_0)^{n-1}$. It is useful to remove the dependence
on the coordinate system by working instead with the invariant 
$F := \pm \sqrt{g_{\alpha\beta} F^\alpha F^\beta} 
= f_0^{-1/2} F^r$, with the sign selected so that 
$\mbox{sign}(F) = \mbox{sign}(F^r)$. This gives 
\begin{equation} 
F = q \partial_r \Phi(r_0, \bm{\Omega}_0), 
\label{eq:SF_formal1} 
\end{equation} 
which represents the magnitude of the force actually measured by a
static observer at $(r_0,\bm{\Omega}_0)$.    
 
\subsection{Decomposition in spherical harmonics} 

To proceed we decompose the potential and charge density in the
higher-dimensional spherical harmonics introduced in the Appendix. 
We write
\begin{equation} 
\Phi(r,\bm{\Omega}) = \sum_{\ell,j} \psi_{\ell,j}(r)
Y_{\ell,j}(\bm{\Omega})
\label{eq:Phi_decomposed} 
\end{equation} 
and 
\begin{equation} 
\delta(\bm{\Omega},\bm{\Omega}_0) = \sum_{\ell,j}
\bar{Y}_{\ell,j}(\bm{\Omega}_0) Y_{\ell,j}(\bm{\Omega}), 
\end{equation} 
where an overbar indicates complex conjugation,
$Y_{\ell,j}(\bm{\Omega})$ are the spherical harmonics, labelled by an
integer degree $\ell$ ($\ell = 0, 1, 2, \cdots$) and a degeneracy
index $j$ that ranges over a number $N(n,\ell)$ of distinct values ---
refer to Eq.~(\ref{eq:degenY}). Making the substitution returns the
sequence of ordinary differential equations  
\begin{equation}  
r^2 \psi''_{\ell,j} + n r \psi'_{\ell,j} 
- \frac{\ell(\ell+n-1)}{f} \psi_{\ell,j} =
\frac{q \Omega_n}{r_0^{n-2}} 
\bar{Y}_{\ell,j}(\bm{\Omega}_0)  \delta(r-r_0) 
\end{equation} 
for the expansion coefficients $\psi_{\ell,j}(r)$; a prime indicates
differentiation with respect to $r$.  

Without loss of generality we may place the particle on the polar
axis. According to Eq.~(\ref{eq:special_value}), this ensures that
only the axisymmetric mode $j=0$ contributes to $\Phi$. Defining 
$\psi_\ell(r) := \sqrt{N(n,\ell)/\Omega_n}\, \psi_{\ell,0}(r)$, we find
that the differential equations become
\begin{equation}  
r^2 \psi''_{\ell} + n r \psi'_{\ell} 
- \frac{\ell(\ell+n-1)}{f} \psi_{\ell} =
\frac{q N(n,\ell)}{r_0^{n-2}}\, \delta(r-r_0).  
\label{eq:deqs} 
\end{equation} 
Making use of Eq.~(\ref{eq:YvsP}), we also find that the scalar
potential can be expressed as 
\begin{equation} 
\Phi(r,\chi) = \sum_\ell \psi_\ell(r) {\cal P}_\ell(\cos\chi), 
\label{eq:potential_decomp}
\end{equation} 
where $\chi := \theta^n$ is the angle from the polar axis, and
${\cal P}_\ell(\cos\chi)$ are the generalized Legendre polynomials
introduced in the Appendix.   

The substitutions
\begin{equation} 
\xi := 2(r/R)^{n-1} - 1, \qquad 
\psi_\ell := \sqrt{f} P(\xi) 
\end{equation} 
bring Eq.~(\ref{eq:deqs}) to the form of an associated Legendre
equation with parameters $\nu = \ell/(n-1)$ and $\mu = 1$. 
The linearly independent solutions to the homogeneous version of
Eq.~(\ref{eq:deqs}) are therefore 
\begin{equation} 
\psi^{\rm in}_\ell = \sqrt{f} P_\nu^1(\xi), \qquad 
\psi^{\rm out}_\ell = \sqrt{f} Q_\nu^1(\xi). 
\label{eq:psi_inner_outer} 
\end{equation}  
The inner solution $\psi^{\rm in}_\ell$ is regular at $\xi=1$ ($r=R$)  
but singular at infinity, while the outer solution 
$\psi^{\rm out}_\ell$ is singular at $\xi=1$ but regular at
infinity. The solution to Eq.~(\ref{eq:deqs}) can be obtained by
combining these solutions and enforcing the appropriate junction
conditions at $r=r_0$. With $\psi^<_\ell$ denoting the solution for 
$r < r_0$, and $\psi^>_\ell$ denoting the solution for $r > r_0$, we
have  
\begin{equation} 
\psi^<_\ell = \frac{qN(n,\ell)}{r_0^n W_\ell} 
\psi^{\rm out}_\ell(r_0) \psi^{\rm in}_\ell(r), \qquad 
\psi^>_\ell = \frac{qN(n,\ell)}{r_0^n W_\ell} 
\psi^{\rm in}_\ell(r_0) \psi^{\rm out}_\ell(r), 
\end{equation} 
where the Wronskian $W_\ell := \psi_{\rm in} \psi'_{\rm out} 
- \psi_{\rm out} \psi'_{\rm in}$ is evaluated at $r=r_0$. Making use
of Eq.~(8.18) of Ref.~\cite{abramowitz-stegun:72}, we find that   
\begin{equation} 
W_\ell = \frac{\ell(\ell+n-1)}{2(n-1)} \frac{R^{n-1}}{r_0^n}. 
\end{equation} 
Our final expression for the solution to Eq.~(\ref{eq:deqs}) is then
\begin{equation} 
\psi^<_\ell = \frac{2(n-1) N(n,\ell)}{\ell(\ell+n-1)} 
\frac{q}{R^{n-1}} \psi^{\rm out}_\ell(r_0) \psi^{\rm in}_\ell(r) 
\label{eq:psi_lesser} 
\end{equation} 
and
\begin{equation} 
\psi^>_\ell = \frac{2(n-1) N(n,\ell)}{\ell(\ell+n-1)} 
\frac{q}{R^{n-1}} \psi^{\rm in}_\ell(r_0) \psi^{\rm out}_\ell(r).  
\label{eq:psi_greater} 
\end{equation} 
Complete expressions can be obtained by inserting
Eq.~(\ref{eq:psi_inner_outer}) with $\nu = \ell/(n-1)$ and 
$\xi = 2(r/R)^{n-1} - 1$.  
 
The special case $\ell = 0$ must be handled separately. Here we find 
\begin{equation} 
\psi^<_0 = -\frac{1}{n-1} \frac{q}{r_0^{n-1}}, \qquad 
\psi^>_0 = -\frac{1}{n-1} \frac{q}{r^{n-1}}. 
\label{eq:psi_zero} 
\end{equation} 
The electric field associated with this solution vanishes for 
$r < r_0$ and is equal to $q/r^n$ for $r > r_0$; these expressions are
compatible with the presence of a charge $q$ at $r=r_0$.  

In terms of the mode decomposition, the (formal) self-force acting on
the charged particle is obtained by inserting
Eq.~(\ref{eq:potential_decomp}) within Eq.~(\ref{eq:SF_formal1})
and setting $r = r_0$, $\chi = 0$. This gives
\begin{equation} 
F = q \sum_\ell \psi'_\ell(r_0)  
\label{eq:SF_formal2} 
\end{equation} 
after making use of the normalization condition ${\cal P}_\ell(1) = 1$
for the generalized Legendre polynomials --- refer to
Eq.~(\ref{eq:Legendre_norm}). Because the electric field is actually
infinite at the position of the charge, this mode sum does not
converge and the computation of the self-force requires
regularization.    
 
\section{Hadamard regularization} 
\label{sec:Hadamard} 

\subsection{Regularization and renormalization} 

We wish to turn Eq.~(\ref{eq:SF_formal2}) into a meaninful
expression for the self-force. We begin by generalizing the context to
a charged particle held at a fixed position $\bm{x}_0$ in any static,
($n+2$)-dimensional spacetime with metric   
\begin{equation} 
ds^2 = -N^2\, dt^2 + h_{ab}\, dx^a dx^b, 
\label{eq:metric_static} 
\end{equation} 
where the lapse $N$ and spatial metric $h_{ab}$ depend on the
$n+1$ spatial coordinates $x^a$ only. The electromagnetic self-force
acting on this particle is expressed formally as 
\begin{equation} 
F^a = q N^{-1} h^{ab} \partial_b \Phi, 
\end{equation} 
in which all quantities are evaluated at $\bm{x}_0$. We wish to
turn this formal statement into something meaningful. 

We assert \cite{quinn-wald:97} that the physical self-force acting on
the particle is 
\begin{equation} 
F^a = q N^{-1} h^{ab}\langle \partial_b \Phi \rangle_{\rm ren}, 
\end{equation} 
in which $\langle \partial_b \Phi \rangle_{\rm ren}$ is the average of
$\partial_b \Phi$ on a small surface $s = \mbox{constant}$ surrounding
the particle, from which all contributions that diverge in the limit
$s \to 0$ are removed. The average is defined precisely by working in
Riemann normal coordinates around $\bm{x}_0$, and $s$ denotes proper
distance from the particle; the averaging is therefore performed on a
surface of constant proper distance. We shall see that the diverging
terms are proportional to the particle's acceleration, so that they
can be absorbed into a redefinition of the particle's mass.  

For a practical implementation of this regularization procedure, it is
convenient to introduce a {\it singular potential} $\Phi^{\sf S}$, a
solution to Maxwell's equations for a point charge at 
$\bm{x}_0$, constructed locally with no regards to boundary
conditions imposed at infinity or anywhere else. The singular
potential is just as singular as $\Phi$ at $\bm{x} = \bm{x}_0$, and
the difference $\Phi - \Phi^{\sf S}$ is smooth. We write 
\begin{equation} 
\langle \partial_b \Phi \rangle_{\rm ren}    
= \partial_b \Phi - \partial_b \Phi^{\sf S} 
+ \langle \partial_b \Phi^{\sf S} \rangle_{\rm ren}, 
\label{eq:regul} 
\end{equation} 
omitting the average sign on the difference 
$\partial_b \Phi - \partial_b \Phi^{\sf S}$ because it is smooth in
the limit $s \to 0$. 

For the next step we return to the specific context of
spherically-symmetric spacetimes, and express the metric in the
general form of 
\begin{equation} 
ds^2 = -e^{2\phi}\, dt^2 + f^{-1}\, dr^2 + r^2\, d\Omega^2_n, 
\label{eq:metric_spherical} 
\end{equation} 
in which $\phi(r)$ and $f(r)$ are arbitrary functions of the radial
coordinate, and $d\Omega^2_n$ is the metric on a unit $n$-sphere. With
the particle placed on the polar axis $\chi = 0$, we decompose $\Phi$
and $\Phi^{\sf S}$ as in Eq.~(\ref{eq:potential_decomp}), and write
the self-force $F := f_0^{-1/2} F^r$ as  
\begin{equation} 
F = F_{\rm mode} + F_{\sf S}, 
\label{eq:F_regmode1} 
\end{equation} 
where 
\begin{equation} 
F_{\rm mode} := q e^{-\phi_0} f_0^{1/2} \sum_\ell 
\bigl[ \psi'_{\ell}(r_0) - \psi^{{\sf S}\prime}_\ell(r_0) \bigr] 
\label{eq:F_regmode2} 
\end{equation}   
is a convergent sum over $\ell$-modes, and 
\begin{equation} 
F_{\sf S} := q e^{-\phi_0} f_0^{1/2} 
\langle \partial_r \Phi^{\sf S} \rangle_{\rm ren} 
\label{eq:F_regmode3} 
\end{equation}  
is the regularized contribution from the singular potential.  We 
introduced the notation $f_0 := f(r_0)$ and $\phi_0 := \phi(r_0)$. 

Our computation of the self-force is based on
Eq.~(\ref{eq:F_regmode1}). We identify the singular potential
$\Phi^{\sf S}$ with the {\it Hadamard Green's function} associated 
with the differential equation satisfied by an electrostatic potential 
$\Phi := \Phi_t$ in a static, $(n+2)$-dimensional spacetime. After
introducing the main equations we review Hadamard's construction in an
arbitrary number of dimensions, and then specialize it to the specific
case of a five-dimensional spacetime ($n=3$). We next construct the
Hadamard Green's function as a local expansion around the base point,
and calculate $\langle \partial_a \Phi^{\sf S} \rangle_{\rm ren}$. 
Then we specialize the results to a spherically-symmetric spacetime,
decompose $\partial_r \Phi^{\sf S}$ in generalized Legendre
polynomials, and calculate the modes 
$\psi^{{\sf S}\prime}_{\ell}(r_0)$ that appear in 
Eq.~(\ref{eq:F_regmode2}); these give rise to the ubiquitous
{\it regularization parameters} of the self-force literature
\cite{barack-ori:00,  barack-etal:02, barack-ori:03a}. This long
computation will return all the ingredients required in the
evaluation of Eq.~(\ref{eq:F_regmode1}).  

\subsection{Green's function in a static spacetime}    

The metric of a static, $(n+2)$-dimensional spacetime is expressed as
in Eq.~(\ref{eq:metric_static}). We introduce the vector field 
\begin{equation} 
A_a := \partial_a \ln N,  
\end{equation} 
and write Maxwell's equation for the potential $\Phi$ as 
\begin{equation} 
\nabla^2 \Phi - A^a \partial_a \Phi = \Omega_n \mu, 
\label{eq:Phi_eq} 
\end{equation} 
where $\mu := N^2 j^t$ and $\nabla^2 := h^{ab} \nabla_a \nabla_b$ is
the Laplacian operator in the $(n+1)$-dimensional space with metric
$h_{ab}$; $\nabla_a$ is the covariant derivative operator in this
space.  

The field equation can be solved by means of a Green's function
$G(\bm{x},\bm{x'})$ that satisfies 
\begin{equation} 
\nabla^2 G(\bm{x},\bm{x'}) - A^a \partial_a G(\bm{x},\bm{x'}) 
= -\Omega_n \delta(\bm{x},\bm{x'}), 
\label{eq:G_eq} 
\end{equation} 
where $\delta(\bm{x},\bm{x'})$ is a scalarized Dirac distribution
defined by $\int a(\bm{x'}) \delta(\bm{x},\bm{x'}) \sqrt{h'}\, 
d^{n+1} x' = a(\bm{x})$ when $\bm{x}$ lies within
the domain of integration; $h'$ is the determinant of the spatial
metric evaluated at $\bm{x'}$, and $a(\bm{x'})$ is an arbitrary test
function of the spatial coordinates. In terms of the Green's function
the solution to Eq.~(\ref{eq:Phi_eq}) is 
\begin{equation} 
\Phi(\bm{x}) = -\int G(\bm{x},\bm{x'}) \mu(\bm{x'}) \sqrt{h'}\,
d^{n+1} x'. 
\label{eq:Phi_G1} 
\end{equation} 
Notice that the source term in the equation for $\Phi$ comes with a
positive sign, while it comes with a negative sign in the equation for
$G$; this difference, which is entirely a matter of convention,
explains the appearance of a negative sign on the right-hand side of
Eq.~(\ref{eq:Phi_G1}). 

For a static charge at a fixed position $\bm{x}_0$, the current
density of Eq.~(\ref{eq:density}) yields $\mu(\bm{x}) = q N(\bm{x}_0) 
\delta(\bm{x},\bm{x}_0)$, and Eq.~(\ref{eq:Phi_G1}) reduces to 
\begin{equation} 
\Phi(\bm{x}) = -q N(\bm{x}_0) G(\bm{x},\bm{x}_0). 
\label{eq:Phi_G2} 
\end{equation} 

\subsection{Hadamard construction} 
\label{subsec:ambiguities} 

The Hadamard Green's function $G_{\rm H}(\bm{x},\bm{x'})$ is a local
solution to Eq.~(\ref{eq:G_eq}) that incorporates the singularity
structure implied by the Dirac distribution, but does not enforce
boundary conditions that we might wish to impose on the potential
$\Phi$ (for example, a falloff condition at spatial infinity). The
theory of such objects was developed by Hadamard (who called them
``elementary solutions'' \cite{hadamard:23}), and it is conveniently
summarized in a number of references \cite{dewitt-brehme:60,
  friedlander:75, poisson-pound-vega:11}. We provide a brief
description of the construction here, but include no derivations.  

The local theory of Green's functions relies heavily on Synge's world
function $\sigma(\bm{x},\bm{x'})$, which is half the
squared geodesic distance between the field point $\bm{x}$ 
and the base point $\bm{x'}$; it is assumed that $\bm{x}$ is
sufficiently close to $\bm{x'}$ that the geodesic joining them is
unique. The gradient of $\sigma$ with respect to $x^a$, denoted
$\sigma_a$, is tangent to the geodesic, and the same is true of
$\sigma_{a'}$, the gradient with respect to $x^{\prime a}$; the
vectors point in opposite directions. The mathematical theory of
two-point tensors (or bitensors), of which $\sigma$, $\sigma_a$, and
$\sigma_{a'}$ are examples, is developed systematically in
Refs.~\cite{synge:60, dewitt-brehme:60} and summarized in
Ref.~\cite{poisson-pound-vega:11}. Our developments below rely heavily
on these techniques.   

The structure of the Hadamard Green's function depends critically on
the dimensionality of the space.  When $n$ is even ($n+1$ odd), the 
Green's function can be expressed as 
\begin{equation} 
G_{\rm H}(\bm{x},\bm{x'}) = \frac{1}{n-1} 
\frac{U (\bm{x},\bm{x'})}{(2\sigma)^{\frac{1}{2}(n-1)}}, 
\label{eq:GH_even} 
\end{equation}
where $U$ is a biscalar that is assumed to be smooth in the
coincidence limit $\bm{x} \to \bm{x'}$. When $n$ is odd ($n+1$ even)
we have instead   
\begin{equation} 
G_{\rm H}(\bm{x},\bm{x'}) = \frac{1}{n-1}
\frac{U (\bm{x},\bm{x'})}{(2\sigma)^{\frac{1}{2}(n-1)}}  
+ V (\bm{x},\bm{x'}) \ln \frac{2\sigma}{\lambda^2} 
+ W (\bm{x},\bm{x'}), 
\label{eq:GH_odd} 
\end{equation} 
where $V$ and $W$ are other smooth biscalars, and $\lambda$ is an
arbitrary length parameter that makes the argument of the logarithm
dimensionless. In both cases $U$ must be normalized by 
$U(\bm{x'},\bm{x'}) = 1$ to ensure that Eq.~(\ref{eq:GH_even})
satisfies Eq.~(\ref{eq:G_eq}).  

When $n$ is even, $U$ is constructed as an expansion in powers of
$2\sigma$,   
\begin{equation} 
U(\bm{x},\bm{x'}) = \sum_{p=0}^\infty U_p(\bm{x},\bm{x'}) 
(2\sigma)^p.  
\label{eq:U_sigma} 
\end{equation} 
Substitution in Eq.~(\ref{eq:GH_even}) and then Eq.~(\ref{eq:G_eq})
reveals that each expansion coefficient must satisfy 
\begin{equation} 
\bigl( 2\sigma^a \nabla_a - A^a \sigma_a + \nabla^2 \sigma 
+ 2p - n - 1 \bigr) U_p = -\frac{2p-n+1}{(n-1)^2} 
\bigl( \nabla^2 - A^a \nabla_a \bigr) U_{p-1}.  
\label{eq:Up_recur} 
\end{equation} 
This is a recursion relation for $U_p$, and the differential operator
$\sigma^a \nabla_a$ on the left-hand side indicates that each equation 
is a transport equation that can be integrated along each geodesic
that emanates from the base point $\bm{x'}$. The equation for $U_0$ is
integrated with a zero right-hand side, and a unique solution is
selected by enforcing the coincidence limit 
$U_0(\bm{x'},\bm{x'}) = 1$. Hadamard proved \cite{hadamard:23} that
the expansion of Eq.~(\ref{eq:U_sigma}) converges in a sufficiently
small neighborhood around $\bm{x'}$.    

When $n$ is odd the construction must be modified to account for the
fact that the right-hand side of Eq.~(\ref{eq:Up_recur}) vanishes when
$p = \frac{1}{2}(n-1)$. The expansion for $U$ must then be truncated
to 
\begin{equation} 
U(\bm{x},\bm{x'}) = \sum_{p=0}^{\frac{1}{2} (n-3)} U_p(\bm{x},\bm{x'}) 
(2\sigma)^p, 
\label{eq:U_sigma_odd} 
\end{equation} 
and the additional terms in Eq.~(\ref{eq:GH_odd}) are inserted to
ensure that the Green's function continues to be a solution to
Eq.~(\ref{eq:G_eq}). The biscalars $V$ and $W$ are also constructed as
expansions in powers of $2\sigma$,  
\begin{equation} 
V(\bm{x},\bm{x'}) = \sum_{p=0}^\infty V_p(\bm{x},\bm{x'}) 
(2\sigma)^p, \qquad 
W(\bm{x},\bm{x'}) = \sum_{p=0}^\infty W_p(\bm{x},\bm{x'}) 
(2\sigma)^p,   
\label{eq:VW_sigma} 
\end{equation} 
and substitution in Eq.~(\ref{eq:GH_odd}) and (\ref{eq:G_eq}) produces
the recursion relations 
\begin{equation} 
\bigl( 2\sigma^a \nabla_a 
- A^a \sigma_a + \nabla^2 \sigma - 2 \bigr) V_0 = 
-\frac{1}{2(n-1)} \bigl( \nabla^2 - A^a \nabla_a \bigr)
U_{\frac{1}{2}(n-3)}, 
\label{eq:V0_recur} 
\end{equation} 
\begin{equation} 
\bigl( 2\sigma^a \nabla_a 
- A^a \sigma_a + \nabla^2 \sigma + 2p - 2 \bigr) V_p 
= -\frac{1}{2p} \bigl( \nabla^2 - A^a \nabla_a \bigr) V_{p-1},   
\label{eq:Vp_recur} 
\end{equation} 
and 
\begin{equation} 
\bigl( 2\sigma^a \nabla_a - A^a \sigma_a + \nabla^2 \sigma 
+ 2p  - 2 \bigr) W_p = -\frac{1}{p} \bigl( 2\sigma^a \nabla_a 
- A^a \sigma_a + \nabla^2 \sigma + 4p  - 2 \bigr) V_p 
- \frac{1}{2p} \bigl( \nabla^2 - A^a \nabla_a \bigr) W_{p-1}.  
\label{eq:Wp_recur} 
\end{equation} 
The recursion relation (\ref{eq:Up_recur}) continues to apply in the
odd case. Equation (\ref{eq:V0_recur}) determines $V_0$ from the last
coefficient in the expansion for $U$, and Eq.~(\ref{eq:Vp_recur})
determines the remaining coefficients $V_p$. Equation
(\ref{eq:Wp_recur}) permits the determination of $W_p$ for  
$p \geq 1$, but there is no equation that determines $W_0$, which must 
remain arbitrary. The expansions of Eq.~(\ref{eq:VW_sigma}) are also
known to converge \cite{hadamard:23} in a sufficiently small
neighborhood around $\bm{x'}$. 

The Hadamard Green's function for $n$ odd is subjected to two types of 
ambiguities. The first concerns the choice of length parameter
$\lambda$, which is arbitrary, and the second concerns the choice of
function $W_0(\bm{x},\bm{x'})$, which is also arbitrary. These
ambiguities are not independent. In fact, the freedom to choose
$\lambda$ is merely a special case of the freedom to choose $W_0$. To
see this, suppose that an initial choice for $W_0$ is shifted to $W_0'  
= W_0 + 2V_0 \ln(\lambda/\lambda')$, where $\lambda'$ is an alternate
choice of length parameter. The shift is then propagated to each $W_p$ by
the recursion relations (\ref{eq:Wp_recur}), and we find that 
$W_p \to W_p' = W_p + 2V_p \ln(\lambda/\lambda')$, which implies that   
\begin{equation} 
W \to W' = W + 2V \ln(\lambda/\lambda'). 
\end{equation} 
This, finally, is equivalent to a shift $\lambda \to \lambda'$ in the
Hadamard form of Eq.~(\ref{eq:GH_odd}). 

\subsection{Local expansion for $n=3$} 
\label{subsec:GH_local} 

We now set $n=3$ and use Eqs.~(\ref{eq:GH_odd}),
(\ref{eq:U_sigma_odd}), (\ref{eq:VW_sigma}) and the recursion
relations of Eqs.~(\ref{eq:Up_recur}), (\ref{eq:V0_recur}),
(\ref{eq:Vp_recur}), (\ref{eq:Wp_recur}) to construct the Hadamard
Green's function as a local expansion about the base point
$\bm{x'}$. To address the ambiguities discussed in the preceding  
paragraph, we specifically set $W_0(\bm{x},\bm{x'})=0$ for some
arbitrary choice of $\lambda$. This choice is justified on the
basis that $W_0$ is a smooth contribution to the Hadamard Green's
function that cancels out when it is incorporated in
Eq.~(\ref{eq:regul}). The remaining terms in the expansion for $W$
play no role in the regularization prescription, because they vanish
in the limit $\bm{x} \to \bm{x'}$.   

Equation (\ref{eq:U_sigma_odd}) indicates that $U = U_0$ when $n=3$,
and Eq.~(\ref{eq:Up_recur}) implies that $U_0$ satisfies the transport
equation 
\begin{equation} 
\bigl( 2\sigma^a \nabla_a - A^a \sigma_a + \nabla^2 \sigma 
- 4 \bigr) U_0 = 0
\label{eq:Uzero_transport} 
\end{equation} 
with $U_0(\bm{x'},\bm{x'}) = 1$. To integrate this equation we
postulate the existence of a local expansion of the form 
\begin{equation} 
U_0 = 1 + a_{a'} \sigma^{a'} 
+ \frac{1}{2} a_{a'b'}  \sigma^{a'} \sigma^{b'} 
+ \frac{1}{6} a_{a'b'c'}  \sigma^{a'} \sigma^{b'} \sigma^{c'} 
+ O(\epsilon^4), 
\label{eq:Uzero_expan} 
\end{equation} 
in which $a_{a'}$, $a_{a'b'}$, and $a_{a'b'c'}$ are tensors defined at
the base point $\bm{x'}$. We let $\epsilon$ be a measure of
distance between $\bm{x}$ and $\bm{x'}$, so that $\sigma^{a'} =
O(\epsilon)$. Noting also that $\sigma = O(\epsilon^2)$, we see that a
truncation of the expansion at order $\epsilon^3$ implies that the
$U_0/(2\sigma)$ contribution to the Green's function is computed
through order $\epsilon$; we shall maintain this degree of accuracy in
the remaining calculations.  

The base-point tensors are determined by inserting
Eq.~(\ref{eq:Uzero_expan}) within Eq.~(\ref{eq:Uzero_transport}) and
solving order by order in $\epsilon$. These manipulations are aided by
the identities $h^{a'}_{\ a} \sigma^a = -\sigma^{a'}$ and
$\sigma^{a'}_{\ ;a} \sigma{^a} = \sigma^{a'}$ satisfied by the world
function, as well as the standard expansions 
\begin{subequations} 
\begin{align}  
\sigma_{;ab} &= h_a^{\ a'} h_b^{\ b'} \biggl[ h_{a'b'} 
- \frac{1}{3} R_{a'c'b'd'} \sigma^{c'} \sigma^{d'} 
+ \frac{1}{4} R_{a'c'b'd';e'} \sigma^{c'} \sigma^{d'} \sigma^{e'} 
+  O(\epsilon^4) \biggr], \\ 
\sigma_{;a'b} &= -h^{b'}_{\ b} \biggl[ h_{a'b'} 
+ \frac{1}{6} R_{a'c'b'd'} \sigma^{c'} \sigma^{d'}
- \frac{1}{12} R_{a'c'b'd';e'} \sigma^{c'} \sigma^{d'} \sigma^{e'} 
+  O(\epsilon^4) \biggr], \\ 
A_a &= h_a^{\ a'} \biggl[ A_{a'} - A_{a';c'} \sigma^{c'} 
+ \frac{1}{2} A_{a';c'd'} \sigma^{c'} \sigma^{d'}     
+  O(\epsilon^3) \biggr], \\ 
h^a_{\ b';c} &= \frac{1}{2} h^a_{\ a'} h^{c'}_{\ c} 
R^{a'}_{\ b'c'd'} \sigma^{d'} + O(\epsilon^2); 
\end{align}
\end{subequations} 
here $h^a_{\ a'}$ is the parallel propagator, which takes a vector
$v^{a'}$ at $\bm{x'}$ and returns the parallel-transported vector 
$v^a = h^a_{\ a'} v^{a'}$ at $\bm{x}$, $R_{a'b'c'd'}$ is the spatial
Riemann tensor (defined with respect to the spatial metric $h_{ab}$)
evaluated at $\bm{x'}$, and a semicolon indicates covariant
differentiation. A straightforward computation returns  
\begin{subequations} 
\label{eq:a_coeffs} 
\begin{align} 
a_{a'} &= -\frac{1}{2} A_{a'}, \\ 
a_{a'b'} &= \frac{1}{2} A_{a';b'} + \frac{1}{4} A_{a'} A_{b'} 
+ \frac{1}{6} R_{a'b'}, \\ 
a_{a'b'c'} &= -\frac{1}{2} A_{(a';b'c')} 
-  \frac{3}{4} A_{(a'} A_{b';c')} 
- \frac{1}{8} A_{a'} A_{b'} A_{c'} 
- \frac{1}{4} A_{(a'} R_{b'c')} 
- \frac{1}{4} R_{(a'b';c')}, 
\end{align} 
\end{subequations} 
where $R_{a'b'}$ is the spatial Ricci tensor at $\bm{x'}$, and indices
enclosed within round brackets are fully symmetrized. While these
calculations were carried out specifically for $n=3$, it is easy to
show that the end result for $U_0$ is actually independent of $n$. 

We next compute the $V\ln(2\sigma)$ contribution to the Green's
function through order $\epsilon$ (formally treating the logarithm as
a quantity of order unity), and this requires $V_0$ expanded through
order $\epsilon$. We write  
\begin{equation} 
V_0 = b + b_{a'} \sigma^{a'} + O(\epsilon^2) 
\label{eq:Vzero_expan} 
\end{equation} 
and determine the coefficients $b$ and $b_{a'}$ by inserting
Eqs.~(\ref{eq:Uzero_expan}) and (\ref{eq:Vzero_expan}) within
Eq.~(\ref{eq:V0_recur}). Another straightforward computation produces 
$b = -\frac{1}{8} a^{a'}_{\ a'} - \frac{1}{8} A^{a'} a_{a'}$ and    
$b_{a'} = -\frac{1}{4} b A_{a'} - \frac{1}{16} a^{b'}_{\ b'a'}
+ \frac{1}{24} a_{b'} R^{b'}_{\ a'} - \frac{1}{16} A^{b'} a_{b'a'} 
+ \frac{1}{16} a_{b'} A^{b'}_{\ ;a'}$, and substitution of
Eqs.~(\ref{eq:a_coeffs}) gives 
\begin{subequations} 
\label{eq:b_coeffs} 
\begin{align} 
b &= -\frac{1}{16} \biggl( A^{a'}_{\ ;a'} 
- \frac{1}{2} A^{a'} A_{a'} + \frac{1}{3} R' \biggr), \\ 
b_{a'} &= \frac{1}{32} \biggl( \nabla^{\prime 2} A_{a'} 
+ A^{b'}_{\ ;b'} A_{a'} - A^{b'} A_{a';b'} 
- \frac{1}{2} A^{b'} A_{b'} A_{a'} 
- A^{b'} R_{a'b'} + \frac{1}{3} R' A_{a'} 
+ \frac{1}{3} R'_{;a'} \biggr), 
\end{align} 
\end{subequations} 
where $R'$ is the Ricci scalar at $\bm{x'}$, and 
$\nabla^{\prime 2} := h^{a'b'} \nabla_{a'} \nabla_{b'}$ is the
Laplacian operator with respect to the variables $\bm{x'}$.  

\subsection{Gradient of the Hadamard Green's function} 

Differentiation of $G_{\rm H}(\bm{x},\bm{x'})$ with respect to $x^a$
yields 
\begin{equation} 
\partial_a G_{\rm H} = -g^{a'}_{\ a} \Biggl\{ 
- \frac{U_0}{(2\sigma)^2} \sigma_{a'} 
+ \frac{1}{2\sigma} \biggl[ \frac{1}{2} a_{a'} 
+ \frac{1}{2} a_{a'b'} \sigma^{b'} 
+ \biggl( \frac{1}{12} a_{e'} R^{e'}_{\ b'a'c'} 
+ \frac{1}{4} a_{a'b'c'} \biggr) \sigma^{b'} \sigma^{c'} 
+ 2 V_0 \sigma_{a'} \biggr] 
+ \ln \frac{2\sigma}{\lambda^2} b_{a'} 
+ O(\epsilon) \Biggr\}, 
\label{eq:gradGH} 
\end{equation}  
and we wish to average this over a surface of constant
proper distance around $\bm{x'}$. For this purpose it is convenient to
introduce Riemann normal coordinates based at $\bm{x'}$, to carry out
all computations in this coordinate system, and to translate back to
the original coordinates when the calculation is completed. The
construction of Riemann normal coordinates is detailed in Sec.~8 of 
Ref.~\cite{poisson-pound-vega:11}. 

The Riemann normal coordinates $x^a$ are intimately tied to Synge's
world function. We have that $\sigma^{a'} = -x^a$, and $2\sigma =
\delta_{ab} x^a x^b := s^2$ is the squared proper distance from the
base point $\bm{x'}$. In these coordinates the parallel propagator is
given explicitly by  
\begin{equation} 
h^{a'}_{\ b} = \delta^a_{\ b} - \frac{1}{6} R^{a'}_{\ c'b'd'} x^c x^d 
- \frac{1}{12} R^{a'}_{\ c'b'd';e'} x^c x^d x^e + O(s^4), 
\end{equation} 
in which the Riemann tensor and its covariant derivatives are
evaluated at $\bm{x'}$. Making the substitutions in
Eq.~(\ref{eq:gradGH}) and writing $x^a = s \Omega^a$, we obtain  
\begin{align} 
\partial_a G_{\rm H} &= -\frac{\Omega_a}{s^3} 
+ \frac{1}{s^2} \biggl( -\frac{1}{2} a_{a'} 
+ a_{b'} \Omega_a \Omega^b \biggr)
+ \frac{1}{s} \biggl( \frac{1}{2} a_{a'b'} \Omega^b 
- \frac{1}{2} a_{b'c'} \Omega_a \Omega^b \Omega^c 
+ 2 b \Omega_a \biggr) 
\nonumber \\ & \quad \mbox{} 
- \frac{1}{4} a_{a'b'c'} \Omega^b \Omega^c 
+ \frac{1}{6} a_{b'c'd'} \Omega_a \Omega^b \Omega^c \Omega^d 
- 2 b_{b'} \Omega_a \Omega^b 
- 2 \ln(s/\lambda) b_{a'} + O(s), 
\end{align} 
in which $\Omega_a := \delta_{ab} \Omega^b$. 

We wish to average $\partial_a G_{\rm H}$ over a surface 
$s = \mbox{constant}$. With $d{\cal A}$ denoting the element of  
surface area, the averaging operation is defined precisely by  
\begin{equation}  
\langle \cdots \rangle 
:= \frac{\int (\cdots)\, d{\cal A}}{\int d{\cal A}}. 
\end{equation} 
The surface is parametrized with polar angles $\theta^A$, and the 
defining relations $x^a = s \Omega^a(\theta^A)$ imply that its
intrinsic metric is given by $ds^2 = s^2 h_{ab} \Omega^a_A
\Omega^b_B\, d\theta^A d\theta^B$, where $\Omega^a_A
:= \partial\Omega^a/\partial\theta^A$. The metric in Riemann
normal coordinates is 
\begin{equation} 
h_{ab} = \delta_{ab} - \frac{1}{3} R_{a'c'b'd'} x^c x^d 
- \frac{1}{6} R_{a'c'b'd';e'} x^c x^d x^e + O(s^4),  
\end{equation} 
and these relations imply that 
\begin{equation} 
d{\cal A} = s^2 \biggl[ 1 - \frac{1}{6} s^2 R_{a'b'} \Omega^a \Omega^b 
- \frac{1}{12} s^3 R_{a'b';c'} \Omega^a \Omega^b \Omega^c 
+ O(s^4) \biggr] d\Omega_n, 
\end{equation} 
where $d\Omega_n$ is the area element on the unit $n$-sphere
introduced in Eq.~(\ref{eq:volume}).  

The explicit expression for $d{\cal A}$ and the standard identities  
\begin{subequations} 
\begin{align} 
\frac{1}{\Omega_n} \int \omega^a\, d\Omega_n &= 0, \\ 
\frac{1}{\Omega_n} \int \omega^a \omega^b\, d\Omega_n 
&= \frac{1}{n+1} \delta^{ab}, \\   
\frac{1}{\Omega_n} \int \omega^a \omega^b \omega^c \, d\Omega_n &= 0,
\\ 
\frac{1}{\Omega_n} \int \omega^a \omega^b \omega^c \omega^d\,
d\Omega_n &= \frac{1}{(n+1)(n+3)} \Bigl( \delta^{ab} \delta^{cd} 
+ \delta^{ac} \delta^{bd}  + \delta^{ad} \delta^{bc} \Bigr) 
\end{align} 
\end{subequations} 
imply that 
\begin{subequations} 
\begin{align} 
\langle \omega^a \rangle &= -\frac{s^3}{6(n+1)(n+3)} \nabla^{a'} R' 
+ O(s^4), \\  
\langle \omega^a \omega^b \rangle &= \frac{1}{n+1} \biggl[ 
\delta^{ab} - \frac{s^2}{3(n+3)} R^{a'b'} 
+ \frac{s^2}{3(n+1)(n+3)} R' \delta^{a'b'} + O(s^4) \biggr], \\ 
\langle \omega^a \omega^b \omega^c \rangle &= O(s^3), \\ 
\langle \omega^a \omega^b \omega^c \omega^d \rangle &= 
\frac{1}{(n+1)(n+3)} \Bigl( \delta^{ab} \delta^{cd} 
+ \delta^{ac} \delta^{bd}  + \delta^{ad} \delta^{bc} \Bigr) 
+ O(s^2). 
\end{align}
\end{subequations} 
With all this we obtain 
\begin{equation} 
\langle \partial_a G_{\rm H} \rangle = -\frac{a_{a'}}{4s^2} 
- \frac{1}{24} a^{b'}_{\ b'a'} - \frac{1}{72} a^{b'} R_{a'b'} 
+ \frac{1}{288} a_{a'} R' + \frac{1}{144} R'_{;a'} 
- \biggl( \frac{1}{2} + 2 \ln \frac{s}{\lambda}\biggr) b_{a'}
+ O(s),
\end{equation} 
after specialization to $n=3$.  

Substitution of Eqs.~(\ref{eq:a_coeffs}) gives 
\begin{equation} 
\langle \partial_a G_{\rm H} \rangle = \frac{A_{a'}}{8 s^3} 
+ c_{a'} - \biggl( \frac{1}{2} + 2 \ln \frac{s}{\lambda} \biggr) b_{a'}
+ O(s), 
\end{equation} 
where 
\begin{equation} 
c_{a'} := \frac{1}{48} \nabla^{\prime 2} A_{a'}  
+ \frac{1}{96} A^{b'}_{\ ;b'} A_{a'} + \frac{1}{48} A^{b'} A_{a';b'} 
+ \frac{1}{192} A^{b'} A_{b'} A_{a'} 
+ \frac{1}{144} A^{b'} R_{a'b'} 
+ \frac{1}{576} R' A_{a'} 
+ \frac{1}{72} R'_{;a'}. 
\label{eq:c_coeffs} 
\end{equation} 
The first term diverges in the limit $s \to 0$, but since the
particle's covariant acceleration at $\bm{x'}$ is precisely equal to
$A_{a'}$, this divergence can be absorbed into a redefinition of the
particle's mass. Removing this term produces 
\begin{equation} 
\langle \partial_a G_{\rm H} \rangle_{\rm ren} = 
c_{a'} - \biggl( \frac{1}{2} + 2 \ln \frac{s}{\lambda} \biggr) b_{a'}, 
\label{eq:gradGH_ren} 
\end{equation} 
and we see that this features a logarithmic dependence upon $s$. This
remaining dependence on the averaging radius cannot be eliminated 
by renormalization. The result of Eq.~(\ref{eq:gradGH_ren}) was 
obtained in Riemann normal coordinates, but since the right-hand side
of the equation is expressed in tensorial form, the result can
immediately be translated to any other coordinate system. It is worth
mentioning that a calculation performed for a four-dimensional
spacetime ($n=2$) would return 
$\langle \partial_a G_{\rm H} \rangle_{\rm ren} = 0$; in this case
$F_{\sf S} = 0$.    
 
\subsection{Spherically-symmetric space} 

We next specialize the results of the preceding subsections to a 
five-dimensional, static and spherically-symmetric spacetime with
metric   
\begin{equation} 
ds^2 = -e^{2\phi}\, dt^2 + f^{-1}\, dr^2 + r^2\, d\Omega^2_3,  
\end{equation} 
in which 
\begin{equation} 
d\Omega^2_3 := d\chi^2 + \sin^2\chi ( d\theta^2 
+ \sin^2\theta\, d\phi^2 ) 
\end{equation} 
is the metric on a unit three-sphere. Placing the base point $\bm{x'}$
on the polar axis $\chi' = 0$ ensures that $G_{\rm H}(\bm{x},\bm{x'})$
is axisymmetric and therefore a function of $r'$, $r$, and $\chi$
only. The covariant local expansion obtained in
Sec.~\ref{subsec:GH_local} can then be turned into an explicit
expansion in powers of $\Delta := r - r'$ and $\chi$. To achieve this
we rely on techniques developed in Sec.~III of 
Ref.~\cite{haas-poisson:06}, which provide an explicit expression for
$\sigma^{a'}$ as an expansion in powers of $w^a := (x-x')^a =
[\Delta,\chi,0,0]$. The Green's function expressed in terms of 
$\Delta$ and $\chi$ is next differentiated with respect to
$\Delta$ to yield $\partial_r G_{\rm H}$.  

For the purposes of eventually expanding $\partial_r G_{\rm H}$ in
generalized Legendre polynomials to implement the mode sum of
Eq.~(\ref{eq:F_regmode2}), it is helpful to replace the dependence on
$\chi$ by a dependence on a function of $\chi$ that is  
well behaved everywhere on the three-sphere. We choose 
$Q := \sqrt{2(1-\cos\chi)}$, and replace each occurrence of $\chi$
by the expansion $\chi = Q + \frac{1}{24} Q^3 + \frac{3}{640} Q^5 
+ O(Q^7)$. After this step we find it convenient to replace the
dependence on $Q$ (which always occurs through $Q^2$) with a
dependence on   
\begin{equation}
\rho^2 := \frac{\Delta^2}{f(r')} + r^{\prime 2} Q^2 
= 2r^{\prime 2} (\delta^2 + 1 - \cos\chi), 
\label{eq:rho_def} 
\end{equation} 
where 
\begin{equation} 
\delta := \frac{|\Delta|}{\sqrt{2 r^{\prime 2} f(r')}}. 
\label{eq:delta_def} 
\end{equation} 
The reason for this substitution is that $\rho^2$ is the leading term
in an expansion of $2\sigma$ in powers of $w^a$; it therefore appears
prominently in our expansion of the Green's function. The
manipulations described in this paragraph are standard fare of
self-force computations, and the methods originate in an
article by Detweiler, Messaritaki, and Whiting
\cite{detweiler-etal:03}. 

The expression we obtain for $\partial_r G_{\rm H}(\Delta,\rho)$ is
much too large to be displayed here. In a schematic notation, we
have a result of the form 
\begin{equation} 
\partial_r G_{\rm H} = (\partial_r G_{\rm H})_{-3}
+ (\partial_r G_{\rm H})_{-2} + (\partial_r G_{\rm H})_{-1}
+ (\partial_r G_{\rm H})_{0} + (\partial_r G_{\rm H})_{\ln} 
+ O(\epsilon), 
\label{eq:GH_r1} 
\end{equation} 
in which a subscript attached to enclosing brackets indicates the
scaling with $\epsilon$. Each term in the sum is given schematically
by 
\begin{subequations} 
\label{eq:GH_r2} 
\begin{align} 
(\partial_r G_{\rm H})_{-3} &= M_{-3} (\Delta/\rho^4), \\ 
(\partial_r G_{\rm H})_{-2} &= M_{-2} (1/\rho^2) + O(\Delta^2/\rho^4) 
+ O(\Delta^4/\rho^6), \\ 
(\partial_r G_{\rm H})_{-1} &= O(\Delta/\rho^2) + O(\Delta^3/\rho^4)
+ O(\Delta^5/\rho^6) + O(\Delta^7/\rho^8), \\ 
(\partial_r G_{\rm H})_{0} &= M_0 + O(\Delta^2/\rho^2) 
+ O(\Delta^4/\rho^4) + O(\Delta^6/\rho^6) + O(\Delta^8/\rho^8)
+ O(\Delta^{10}/\rho^{10}), \\ 
(\partial_r G_{\rm H})_{\ln} &= M_{\ln} \ln(\rho/\lambda), 
\end{align}
\end{subequations} 
where the coefficients in front of the factors $\Delta^p/\rho^q$
are functions of $r'$ only. The terms involving $M_{-3}$, $M_{-2}$,
$M_0$, and $M_{\ln}$ are those that give rise to regularization
parameters for self-force computations; the remaining terms are
unimportant for our purposes.  

\subsection{Decomposition in Legendre polynomials} 

We next submit $\partial_r G_{\rm H}$ to a decomposition in the
generalized Legendre polynomials introduced in the Appendix. We write  
\begin{equation} 
\partial_r G_{\rm H} = \sum_\ell 
(\partial_r G_{\rm H})_\ell {\cal P}_\ell(\cos\chi), 
\label{eq:GH_decomp}
\end{equation} 
where the expansion coefficients $(\partial_r G_{\rm H})_\ell$ depend 
on $\Delta$ only. For the purposes of evaluating the mode sum of
Eq.~(\ref{eq:F_regmode2}), it is sufficient to obtain  
$(\partial_r G_{\rm H})_\ell$ in the limit $\Delta \to 0$.

For $n=3$ the generalized Legendre polynomials ${\cal P}_\ell(u)$ are
directly related to $U_\ell(u)$, the Chebyshev polynomials of the
second kind. The relation is obtained by comparing their generating
functions [Eq.~(22.9.10) of Ref.~\cite{abramowitz-stegun:72} versus 
Eq.~(\ref{eq:Legendre_generating})], and we obtain  
$U_\ell(u) = (\ell+1) P_\ell(u)$. This allows us to write the
Rodrigues formula [Eq.~(22.11.4) of Ref.~\cite{abramowitz-stegun:72}]   
\begin{equation} 
{\cal P}_\ell(u) = \frac{(-1)^\ell}{(2\ell+1)!!} (1-u^2)^{-1/2} 
\frac{d^\ell}{du^\ell} (1-u^2)^{\ell+\frac{1}{2}} 
\label{eq:Pell_Rodrigues} 
\end{equation} 
and state the orthonormality property [Eq.~(22.2.5) of
Ref.~\cite{abramowitz-stegun:72}] 
\begin{equation} 
\int_{-1}^1 (1-u^2)^{1/2} {\cal P}_\ell(u) {\cal P}_{\ell'}(u)\, du 
= \frac{\pi}{2} \frac{1}{(\ell+1)^2} \delta_{\ell\ell'}.
\label{eq:Pell_ortho} 
\end{equation} 
It follows from this that any function $f(u)$ can be decomposed as   
\begin{equation} 
f(u) = \sum_\ell f_\ell\, {\cal P}_\ell(u), 
\label{eq:f_expanded1} 
\end{equation} 
with coefficients 
\begin{equation} 
f_\ell = \frac{2}{\pi} (\ell+1)^2 \int_{-1}^1 (1-u^2)^{1/2} f(u) 
{\cal P}_\ell(u)\, du. 
\label{eq:f_expanded2}
\end{equation} 
Insertion of Eq.~(\ref{eq:Pell_Rodrigues}) and integration by parts
yields the alternative expression 
\begin{equation} 
f_\ell = \frac{2}{\pi} \frac{(\ell+1)^2}{(2\ell+1)!!} \int_{-1}^1 
(1-u^2)^{\ell+\frac{1}{2}} \frac{d^\ell f}{d u^\ell}\, du 
\label{eq:f_expanded3}
\end{equation} 
for the expansion coefficients. 

The decomposition of $\partial_r G_{\rm H}$, as expressed in
Eqs.~(\ref{eq:GH_r1}) and (\ref{eq:GH_r2}), requires the decomposition
of $\rho^{-p}$, with $\rho$ defined by Eq.~(\ref{eq:rho_def}), and
with $p$ ranging from 2 to 10. To accomplish this we adapt the method
devised by Detweiler, Messaritaki, and Whiting
\cite{detweiler-etal:03}. We set $t = e^{-T}$ in the generating
function of Eq.~(\ref{eq:Legendre_generating}) to obtain    
\begin{equation} 
(\cosh T - u)^{-1} = \sum_\ell 2(\ell+1) {\cal P}_\ell(u) 
e^{-(\ell+1) T}, 
\end{equation} 
write $u = \cos\chi$, $\cosh T = \delta^2 + 1$, and expand the
right-hand side in powers of $\delta$. This yields  
\begin{equation} 
(\delta^2 + 1 - \cos\chi)^{-1} = \sum_\ell A^1_\ell(\delta)\, 
{\cal P}_\ell(\cos\chi) 
\end{equation} 
with 
\begin{equation} 
A^1_\ell = 2(\ell+1) - 2\sqrt{2} (\ell+1)^2 \delta
+ O(\delta^2). 
\end{equation} 
This identity can be used to decompose $\rho^{-2}$, and for the higher 
powers we generate additional identities by repeatedly differentiating
each side with respect to $\delta$. We thus obtain 
\begin{equation} 
(\delta^2 + 1 - \cos\chi)^{-q} = \sum_\ell A^q_\ell(\delta)\, 
{\cal P}_\ell(\cos\chi) 
\end{equation} 
with 
\begin{subequations}
\begin{align}  
A^2_\ell &= \sqrt{2} (\ell+1)^2 \delta^{-1} + O(1), \\
A^3_\ell &= \frac{\sqrt{2}}{4} (\ell+1)^2 \delta^{-3} 
+ O(\delta^{-1}), 
\end{align} 
\end{subequations} 
$A^4_\ell = O(\delta^{-5})$, and $A^5_\ell = O(\delta^{-7})$. 
Combining these results with Eqs.~(\ref{eq:rho_def}) and
(\ref{eq:delta_def}), we arrive at the decompositions 
\begin{equation} 
\rho^{-p} = \sum_\ell (\rho^{-p})_\ell\, {\cal P}_\ell(\cos\chi) 
\label{eq:rhop_decomp1} 
\end{equation} 
with 
\begin{subequations} 
\label{eq:rhop_decomp2} 
\begin{align} 
(\rho^{-2})_\ell &= \frac{\ell+1}{r^{\prime 2}} + O(\Delta), \\ 
(\rho^{-4})_\ell &=  (\ell+1)^2 \frac{\sqrt{f(r')}}{2 r^{\prime 3}} 
\frac{1}{|\Delta|} + O(1), \\ 
\end{align} 
\end{subequations} 
$(\rho^{-6})_\ell = O(\Delta^{-3})$,  
$(\rho^{-8})_\ell = O(\Delta^{-5})$, and  
$(\rho^{-10})_\ell = O(\Delta^{-7})$.  
 
To complete the decomposition of $\partial_r G_{\rm H}$ we must next
obtain a decomposition for $\ln(\rho/\lambda)$. We have 
\begin{equation} 
\ln(\rho/\lambda) = \ln(\sqrt{2} r'/\lambda) 
+ \frac{1}{2} \ln(\delta^2+1-\cos\chi), 
\end{equation} 
and the last term can be decomposed by inserting 
$f(u) := \ln(\delta^2+1-u)$ within Eq.~(\ref{eq:f_expanded3}). The
coefficients $f_\ell(\delta^2)$ can be expanded in powers of
$\delta^2$, and the leading-order term is given by 
\begin{equation} 
f_\ell(0) = -\frac{2}{\pi} \frac{(\ell-1)!(\ell+1)^2}{(2\ell+1)!!} 
\int_{-1}^1 (1-u)^{1/2} (1+u)^{\ell+\frac{1}{2}}\, du
\end{equation} 
when $\ell \neq 0$. The substitution $u = 1-2t$ brings
the integral to the standard form of a Beta function, and after
simplification we obtain  
\begin{equation} 
f_\ell(\delta^2) = -\frac{2(\ell+1)}{\ell(\ell+2)} + O(\delta^2).  
\end{equation} 
For $\ell = 0$ we may insert $f(u)$ directly in
Eq.~(\ref{eq:f_expanded2}) and evaluate the integral. This yields 
$f_0(\delta^2) = \frac{1}{2} - \ln 2 + O(\delta^2)$. Collecting results,
we have established that the decomposition of $\ln(\rho/\lambda)$ in
generalized Legendre polynomials comes with the coefficients  
\begin{subequations} 
\label{eq:logrho_decomp} 
\begin{align} 
\bigl( \ln(\rho/\lambda) \bigr)_{0} &= \ln(r'/\lambda) 
+ 1/4 + O(\Delta^2), \\ 
\bigl( \ln(\rho/\lambda) \bigr)_{\ell \neq 0} &= 
-\frac{\ell+1}{\ell(\ell+2)} + O(\Delta^2). 
\end{align} 
\end{subequations} 

\subsection{Regularization parameters} 

Inserting the decompositions of Eqs.~(\ref{eq:rhop_decomp2}) and
(\ref{eq:logrho_decomp}) into Eqs.~(\ref{eq:GH_r1}) and
(\ref{eq:GH_r2}) produces Eq.~(\ref{eq:GH_decomp}) with the
coefficients  
\begin{subequations}
\begin{align} 
(\partial_r G_{\rm H})_0 &= \frac{\sqrt{f(r')} M_{-3}}{2 r^{\prime 3}}
\mbox{sign}(\Delta) + \frac{M_{-2}}{r^{\prime 2}} + M_0 
+ M_{\ln} \bigl[ \ln(r'/\lambda) + 1/4 \bigr] + O(\Delta),
\\ 
(\partial_r G_{\rm H})_{\ell\neq 0} &= 
\frac{\sqrt{f(r')} M_{-3}}{2 r^{\prime 3}} 
\mbox{sign}(\Delta) (\ell+1)^2 
+ \frac{M_{-2}}{r^{\prime 2}} (\ell+1)  
- M_{\ln} \frac{\ell+1}{\ell(\ell+2)} + O(\Delta). 
\end{align} 
\end{subequations} 
Taking into account Eq.~(\ref{eq:Phi_G2}), $\Phi^{\sf S}$ is
identified with $-q N(\bm{x}_0) G_{\rm H}(\bm{x},\bm{x}_0)$, and its
decomposition involves the coefficients 
$-q e^{\phi_0} (\partial_r G_{\rm H})_\ell$, where
$\phi_0 := \phi(r_0)$. Substituting the previous results gives 
\begin{subequations} 
\label{eq:regparam1} 
\begin{align} 
q^{-1} \psi^{{\sf S}\prime}_0(r_0) &= A + B + C 
+ D \bigl[ \ln(r_0/\lambda) + 1/4 \bigr], \\
q^{-1} \psi^{{\sf S}\prime}_{\ell\neq 0}(r_0) &= A(\ell+1)^2 
+ B(\ell+1) - D \frac{\ell+1}{\ell(\ell+2)},  
\end{align} 
\end{subequations} 
where 
\begin{subequations} 
\begin{align} 
A &:= -\frac{e^{\phi_0} \sqrt{f_0} M_{-3}}{2r_0^3} 
\mbox{sign}(\Delta), \\ 
B &:= -\frac{e^{\phi_0} M_{-2}}{r_0^2}, \\ 
C &:= -e^{\phi_0} M_0, \\ 
D &:= -e^{\phi_0} M_{\ln} 
\end{align} 
\end{subequations}    
are the so-called regularization parameters. We have set 
$f_0 := f(r_0)$. 

Expressions for the regularization parameters can be obtained 
by inserting the explicit form of the coefficients $M_{-3}$, $M_{-2}$,
$M_0$, and $M_{\ln}$, which were left implicit in
Eq.~(\ref{eq:GH_r2}). We find   
\begin{subequations} 
\label{eq:regparam2} 
\begin{align} 
A &= \frac{e^{\phi_0}}{2 r_0^3 \sqrt{f_0}}, \\ 
B &= \frac{e^{\phi_0}}{4 r_0^3} \bigl(2 - r_0 \phi_0' \bigr), \\ 
C &= \frac{e^{\phi_0}}{96 r_0^3} \Bigl[ 
2 \bigl( 8 - r_0 \phi_0' \bigr)  
- \bigl( 16 - 20 r_0 \phi'_0 + 3 r_0^2 \phi^{\prime 2}_0 
- 6 r_0^2 \phi''_0 \bigr) f_0 
- \bigl( 5 - 4 r_0 \phi'_0 \bigr) r_0 f'_0 
+ r_0^2 f''_0 \Bigr], \\ 
D &= -\frac{e^{\phi_0}}{32 r_0^3} \Bigl[ 
4 \bigl( 2 - r_0 \phi_0' \bigr)  
- \bigl( 8 - 10 r_0 \phi'_0 + 6 r_0^2 \phi^{\prime 2}_0 
+ 6 r_0^2 \phi''_0 - r_0^3 \phi^{\prime 3}_0 
+ 2 r_0^3 \phi'''_0 \bigr) f_0 
\nonumber \\ & \quad \mbox{} 
+ \bigl( 2 - 4 r_0 \phi'_0 - 3 r_0^2 \phi''_0 \bigr) r_0 f'_0 
+ \bigl( 2 - r_0 \phi'_0 \bigr) r_0^2 f''_0 \Bigr],  
\end{align} 
\end{subequations}
where $\phi'_0 := \phi'(r_0)$, $\phi''_0 := \phi''(r_0)$, with a
similar notation extending to $f'_0$, $f''_0$, and higher
derivatives. In the case of the Schwarzschild-Tangherlini spacetime,
$e^{2\phi} = f = 1 - (R/r)^2$, and the regularization parameters
become 
\begin{subequations} 
\label{eq:regparam3} 
\begin{align} 
A &= \frac{1}{2r_0^3}\, \mbox{sign}(\Delta), \\ 
B &= \frac{2 - 3 R^2/r_0^2}{4 r_0^3 f_0^{1/2}}, \\ 
C &= -\frac{3 R^4}{32 r_0^7 f_0^{1/2}}, \\ 
D &= \frac{3(6-5R^2/r_0^2)R^4}{32 r_0^7 f_0^{3/2}}. 
\end{align} 
\end{subequations} 

With the regularization parameters now in hand, the mode
sum of Eq.~(\ref{eq:F_regmode2}) becomes 
\begin{equation} 
F_{\rm mode} = q^2 \sum_{\ell} F^{\rm mode}_\ell 
\label{eq:SF_reg3}
\end{equation} 
with 
\begin{subequations} 
\label{eq:SF_reg4} 
\begin{align} 
F^{\rm mode}_0 &= q^{-1} \psi'_0(r_0) - A - B - C 
- D \bigl[ \ln(r_0/\lambda) + 1/4 \bigr], \\ 
F^{\rm mode}_{\ell \neq 0} &= q^{-1} \psi'_{\ell \neq 0}(r_0) 
- A(\ell+1)^2 - B(\ell+1) + D \frac{\ell+1}{\ell(\ell+2)}. 
\end{align} 
\end{subequations}   
To this we must add the contribution $F_{\sf S}$ from the singular
potential, which can be obtained from
Eq.~(\ref{eq:gradGH_ren}). Because $F_{\sf S}$ is spherically
symmetric, it has the effect of modifying the expression for the zero
mode $F_0$. Noting that $D = 2 e^{\phi_0} b_r$ and letting   
\begin{align} 
E &:= e^{\phi_0} c_r \nonumber \\ 
&= -\frac{e^{\phi_0}}{192 r_0^3} \Bigl[ 
2 \bigl( 16 - r_0 \phi_0' \bigr)  
- \bigl( 32 - 14 r_0 \phi'_0 + 6 r_0^2 \phi^{\prime 2}_0 
+ 12 r_0^2 \phi''_0 + r_0^3 \phi^{\prime 3}_0 
+ 6 r_0^3 \phi'_0 \phi''_0 
+ 4 r_0^3 \phi'''_0 \bigr) f_0 
\nonumber \\ & \quad \mbox{} 
+ \bigl( 8 - 3 r_0 \phi'_0 -3 r_0^2 \phi^{\prime 2}_0  
- 6 r_0^2 \phi''_0 \bigr) r_0 f'_0 
+ 2 \bigl( 4 - r_0 \phi'_0 \bigr) r_0^2 f''_0 \Bigr] 
\nonumber \\ 
&= -\frac{(4-7R^2/r_0^2)R^4}{192 r_0^7 f_0^{3/2}},   
\label{eq:regparam4} 
\end{align} 
we find that 
\begin{equation} 
F = F_{\rm mode} + F_{\sf S} = q^2 \sum_{\ell} F_\ell 
\label{eq:SF_reg5} 
\end{equation} 
with 
\begin{subequations} 
\label{eq:SF_reg6} 
\begin{align} 
F_0 &= q^{-1} \psi'_0(r_0) - A - B - C 
- D \ln(r_0/s) - E, \\ 
F_{\ell \neq 0} &= q^{-1} \psi'_{\ell \neq 0}(r_0) 
- A(\ell+1)^2 - B(\ell+1) + D \frac{\ell+1}{\ell(\ell+2)}. 
\end{align} 
\end{subequations}   
At this final stage we notice that the dependence on the arbitrary
length parameter $\lambda$ has completely disappeared. The self-force,
however, retains a dependence on the small averaging radius $s$
introduced in the calculation of $F_{\sf S}$.     
  
The task of providing a regularization prescription for the mode-sum
computation of the self-force is now completed. Our methods, based on 
an identification of the singular potential $\Phi^{\sf S}$ with the
Hadamard Green's function, can be applied to any static,
spherically-symmetric spacetime in five dimensions. 

\section{Evaluation of the self-force} 
\label{sec:SF} 

\subsection{Numerical evaluation}
\label{subsec:numer} 

The electromagnetic self-force acting on a particle of charge $q$
held at position $r_0$ in the five-dimensional
Schwarzschild-Tangherlini spacetime is computed by involving the modes  
of the scalar potential displayed in
Eqs.~(\ref{eq:psi_inner_outer}), (\ref{eq:psi_lesser}),
(\ref{eq:psi_greater}), (\ref{eq:psi_zero}), as well as the
regularization parameters of Eqs.~(\ref{eq:regparam3}) and
(\ref{eq:regparam4}), in the mode sum of Eq.~(\ref{eq:SF_reg5}). The
modes can be evaluated either at $r = r_0^+$, in which case 
$\Delta = 0^+$, or they can be evaluated at $r = r_0^-$, in which case
$\Delta = 0^-$; the notation $a^\pm$ signifies the limit when
$\epsilon \to 0$ of $a \pm \epsilon$, with $\epsilon$ taken to be
strictly positive. The associated Legendre functions are evaluated to
arbitrary numerical accuracy with the symbolic manipulation software
{\tt Maple}, and the mode sum is truncated after a sufficient number
of terms to ensure convergence. The computation requires making a
choice of averaging radius $s$, and for illustrative purposes we
sample the values $s/R = \{ 10^{-4}, 10^{-6}, 10^{-8} \}$.  

The results of this computation were displayed in
Fig.~\ref{fig:emsf1}. The numerical data reveal that the self-force  
$F$ approaches the (positive) asymptotic value $F_{\rm asymp} 
= q^2 R^2/(2r_0^5)$ when $r_0 \gg R$, turns negative when $r$ 
becomes comparable to $4R$, and diverges as $-f_0^{-3/2}$ when 
$r_0 \to R$; here $f_0 := 1 - (r_0/R)^2$.   

\subsection{Large-$r$ expansion of the self-force} 

To gain insight into the large-$r_0$ behavior of the self-force, we
submit the modes of Eqs.~(\ref{eq:psi_inner_outer}) and the
regularization parameters of Eqs.~(\ref{eq:regparam3}) and
(\ref{eq:regparam4}) to an expansion in powers of $R/r_0$, and
evaluate the mode sum analytically. In the remainder of this section
we shall write $r_0 = r$ to simplify the notation; there is no longer
a need to keep these quantities distinct. 

The large-$r$ behavior of $\psi^{\rm in}_\ell$, as defined in
Eq.~(\ref{eq:psi_inner_outer}), can be extracted by making use of
Eq.~(8.772.3) of Ref.~\cite{gradshteyn-ryzhik:80} and Eq.~(15.3.11) of 
Ref.~\cite{abramowitz-stegun:72}. We find that the mode 
function can be expressed as 
\begin{align} 
\psi^{\rm in}_\ell &= 
\frac{\Gamma(\ell+1)}{\Gamma(\ell/2+1) \Gamma(\ell/2)}   
(r/R)^\ell \biggl\{ 
\sum_{p=0}^\ell \frac{ (-\ell/2)_p (-\ell/2-1)_p}{p! (-\ell)_p}
(R/r)^{2p} 
+ \frac{(-1)^\ell \Gamma(\ell/2+1) \Gamma(\ell/2)} 
{\Gamma(\ell+1) \Gamma(-\ell/2) \Gamma(-\ell/2-1)} 
(R/r)^{2\ell+2} 
\nonumber \\ & \quad \mbox{} \times 
\sum_{p=0}^\infty 
\frac{(\ell/2+1)_p (\ell/2)_p}{p!(p+\ell+1)!} (R/r)^{2p} 
\bigl[ -2\ln(r/R) - \psi(p+1) - \psi(p+\ell+2)
+ \psi(\ell/2+p+1) + \psi(\ell/2+p) \bigr] \biggr\}, 
\label{eq:psiin_large_r} 
\end{align} 
in which $(a)_p := a(a+1)\cdots(a+p-1)$ is the Pochhammer symbol, and
$\psi(a)$ is the Digamma function. The mode function is expressed as
an expansion in powers of $R/r$, and the second sum over $p$ can be
truncated when the required degree of accuracy is achieved; this sum
is multiplied by a vanishing factor when $\ell$ is even.  
    
The large-$r$ behavior of $\psi^{\rm out}_\ell$ can be extracted by
making use of Eq.~(8.703) of Ref.~\cite{gradshteyn-ryzhik:80} and
Eq.~(15.3.20) of Ref.~\cite{abramowitz-stegun:72}. Here we find that
the mode function can be expressed as 
\begin{equation} 
\psi^{\rm out}_\ell = 
-\frac{\sqrt{\pi}\, \Gamma(\ell/2+2)}{2^{\ell+2} \Gamma(\ell/2+3/2)} 
(1 - R^2/r^2) (R/r)^{\ell+2} F(\ell/2+1, \ell/2+2;\ell+2;R^2/r^2), 
\label{eq:psiout_large_r} 
\end{equation} 
where $F(a,b;c;z)$ is the hypergeometric function, which is defined as
an infinite expansion in powers of its argument; this expansion also
can be truncated to the desired degree of accuracy.  

The calculation proceeds by first selecting $O$, the maximum power of 
$R/r$ (beyond the leading order) that one wishes to keep in all
expressions. The selected value of $O$ dictates at which point the
infinite sums over $p$ can be truncated in
Eqs.~(\ref{eq:psiin_large_r}) and (\ref{eq:psiout_large_r}). It also
determines $\ell_{\rm max} := \frac{1}{2} O - 1$, the maximum value of
$\ell$ beyond which the second set of terms in
Eq.~(\ref{eq:psiout_large_r}) --- those involving the logarithm and
Digamma functions --- are no longer required in 
$\psi^{\rm in}_\ell$. The mode functions are evaluated
individually for $0 \leq \ell \leq \ell_{\rm max}$, and then inserted
in Eq.~(\ref{eq:psi_lesser}) or (\ref{eq:psi_greater}) to be
substituted in the regularized mode sum of Eq.~(\ref{eq:SF_reg5}). The
partial sum up to $\ell_{\rm max}$ is regularized by inserting the
regularization parameters expanded in powers of $R/r$. Finally, the
remaining sum from $\ell_{\rm max}$ to $\infty$, involving the
expanded mode functions and regularization parameters, is evaluated
exactly in closed form, and added to the partial sum.       

This calculation produces an explicit expression for the self-force,
given as an expansion in powers of $R/r$ truncated to the selected
order $O$. We get 
\begin{equation} 
F = \frac{q^2 R^2}{2 r^5} \biggl( F_{\rm poly} 
- \frac{9}{8} \frac{R^2}{r^2} F_{\ln}\, \ln \frac{4 r^2}{s R} \biggr), 
\label{eq:F_large_r} 
\end{equation} 
where 
\begin{align} 
F_{\rm poly} &=
1 + {\frac {83}{48}}\, x + {\frac {137}{96}}\, {x}^{2}
+ {\frac {741}{512}}\, {x}^{3} + {\frac {2333}{1536}}\,{x}^{4}
+ {\frac {39487}{24576}}\,{x}^{5}
+ {\frac {34827}{20480}}\,{x}^{6}
+ {\frac {1177231}{655360}}\,{x}^{7}
+ {\frac {1736163}{917504}}\,{x}^{8}
+ {\frac {58354439}{29360128}}\,{x}^{9}
\nonumber \\ & \quad \mbox{} 
+ {\frac {91678535}{44040192}}\,{x}^{10}
+ {\frac {2553794009}{1174405120}}\,{x}^{11}
+ {\frac {4181839333}{1845493760}}\,{x}^{12}
+ {\frac {41740066799}{17716740096}}\,{x}^{13}
+ {\frac {46918332385}{19193135104}}\,{x}^{14}
\nonumber \\ & \quad \mbox{} 
+ {\frac {21768997188375}{8598524526592}}\,{x}^{15}
+ {\frac {22506963123633}{8598524526592}}\,{x}^{16}
+ {\frac {1486958228186313}{550305569701888}}\,{x}^{17}
+ O(x^{18}) 
\label{eq:Fpoly1} 
\end{align} 
and 
\begin{align} 
F_{\ln} &= 1 + \frac{2}{3}\,x + \frac{5}{8}\,{x}^{2}
+ \frac{5}{8}\,{x}^{3} + {\frac {245}{384}}\,{x}^{4}
+ {\frac {21}{32}}\,{x}^{5} + {\frac {693}{1024}}\,{x}^{6}
+ {\frac {715}{1024}}\,{x}^{7} + {\frac {23595}{32768}}\,{x}^{8}
+ {\frac {12155}{16384}}\,{x}^{9}
+ {\frac {600457}{786432}}\,{x}^{10}
\nonumber \\ & \quad \mbox{} 
+ {\frac {205751}{262144}}\,{x}^{11} 
+ {\frac {3380195}{4194304}}\,{x}^{12} 
+ {\frac {1300075}{1572864}}\,{x}^{13}
+ {\frac {28415925}{33554432}}\,{x}^{14}
+ {\frac {29084535}{33554432}}\,{x}^{15}
+{\frac {1903421235}{2147483648}}\,{x}^{16} 
+ O(x^{17}), 
\label{eq:Fln1} 
\end{align} 
with $x := (R/r)^2$. These expressions were obtained by selecting 
$O = 36$ and $\ell_{\rm max} = 17$.

\subsection{Summing the large-$r$ expansion} 

The large-$r$ expansion of Eq.~(\ref{eq:F_large_r}), carried out to
such a high order in $R/r$, can be shown to produce an excellent fit
to the numerical data presented in Sec.~\ref{subsec:numer}. The fit,
however, becomes relatively poor as $R/r \to 1$, and indeed, one can
see from Eqs.~(\ref{eq:Fpoly1}) and (\ref{eq:Fln1}) that the expansions
may not converge in this limit. In an effort to produce better fits,
we determined from the numerical data that the self-force appears to
diverge as $f^{-3/2}$ when $r \to R$; we recall that $f = 1-R^2/r^2 
= 1-x$. Removing this factor from the expansions produces 
\begin{align} 
f^{3/2} F_{\rm poly} &= 1 + {\frac {11}{48}}\,x 
- {\frac {19}{24}}\,{x}^{2} + {\frac {9}{512}}\,{x}^{3}
+ {\frac {15}{1024}}\,{x}^{4} + {\frac {103}{8192}}\,{x}^{5}
+ {\frac {903}{81920}}\,{x}^{6}
+ {\frac {1287}{131072}}\,{x}^{7}
+ {\frac {16251}{1835008}}\,{x}^{8} 
+ {\frac {236871}{29360128}}\,{x}^{9}
\nonumber \\ & \quad \mbox{} 
+ {\frac {62161}{8388608}}\,{x}^{10} 
+ {\frac {344923}{50331648}}\,{x}^{11}
+ {\frac {2352909}{369098752}}\,{x}^{12}
+ {\frac {17597487}{2952790016}}\,{x}^{13}
+ {\frac {39054973}{6979321856}}\,{x}^{14}
\nonumber \\ & \quad \mbox{} 
+ {\frac {4122846057}{781684047872}}\,{x}^{15}
+ {\frac {2999320659}{601295421440}}\,{x}^{16}
+ {\frac {2601240413}{549755813888}}\,{x}^{17}
+ O(x^{18}) 
\label{eq:Fpoly2} 
\end{align} 
and 
\begin{equation} 
f^{3/2} F_{\ln} = 1 - \frac{5}{6}\, x + O(x^{17}). 
\label{eq:Fln2} 
\end{equation} 
Remarkably, the factorization allows us to express $F_{\ln}$ in
closed form (assuming that the pattern identified up to order $x^{17}$ 
is not broken at higher orders), and produces what appears to be a
converging series for $f^{3/2} P_{\rm poly}$. This new representation
of the self-force can be shown to give rise to a perfect fit to the
numerical data.  

Our unexpected success at expressing $F_{\ln}$ in closed form
motivated us to seek a means to sum the power expansion of
Eq.~(\ref{eq:Fpoly2}). Let $g_0$ stand for $f^{-3/2} F_{\rm poly}$,
and let $s_0(p)$ be the associated sequence of coefficients, so that
$g_0 = \sum_{p=1}^\infty s_0(p) x^{p-1}$. To search for a pattern in
the sequence we factorize each member in its prime factors, and notice
that the denominators are mostly powers of 2, except for some factors
that can be removed by defining a new sequence\footnote{Multiplication 
  by $p-1$ is optional.} $s_1(p) := (p-1)(p-2)(p-3) s_0(p)$. The new
sequence loses track of the first three members of the original
sequence, but these will be reinstated at a later stage. To proceed
with the search we repeatedly take differences between adjacent
members by defining the new sequences 
$s_2(p) := s_1(p+1) - s_1(p)$, $s_3(p) := s_2(p+1) - s_2(p)$, and 
$s_4(p) := s_3(p+1) - s_3(p)$. At this point a recognizable pattern
reveals itself, and we find that  
\begin{equation} 
s_4(p) = \frac{9(p+8)(2p-1)!!}{2^{p+4} (p+3)!}. 
\end{equation} 
Remarkably, the associated function $g_4 = \sum_{p=1}^\infty s_4(p) 
x^{p-1}$ can be written in closed form:  
\begin{equation} 
g_4 = -\frac{3}{16 x^4} (4 x^3 - 21 x^2 + 24 x - 8) 
- \frac{3}{4 x^4} (2-3x)(1-x)^{3/2}. 
\end{equation} 
From this point on it is a simple matter to reconstruct the function 
$g_0$. Each sequence difference amounts to a
multiplication by $(1-x)/x$, so that $g_1 = x^3 g_4/(1-x)^3$ is the
function associated with $s_1(p)$. To obtain $g_0$ we must account
for the denominator in the relation $s_0(p) =
s_1(p)/[(p-1)(p-2)(p-3)]$. It is easy to see that the divisions by
$p-1$, $p-2$, and then $p-3$ give rise to the following sequence of
operations on $g_1$:  
\begin{subequations} 
\label{eq:g_integrals} 
\begin{align} 
g_1^{p-1} &:= \int_0^x x^{\prime -1} g_1(x')\, dx', \\
g_1^{p-2} &:= \int_0^x x^{\prime -2} g_1^{p-1}(x')\, dx', \\ 
g_1^{p-3} &:= \int_0^x x^{\prime -2} g_1^{p-2}(x')\, dx'. 
\end{align} 
\end{subequations} 
Reinstating the early terms that were erased when $s_1(p)$ was
introduced, we finally obtain 
\begin{equation} 
g_0 = 1 + \frac{11}{48}\, x - \frac{19}{24}\, x^2 + x^2 g_1^{p-3}. 
\end{equation}
The integrations displayed in Eq.~(\ref{eq:g_integrals}) are
elementary, and we finally arrive at 
\begin{equation} 
f^{3/2} F_{\rm poly} = 
- \frac{1}{4x} + \frac{5}{8} + \frac{139}{96}\, x - \frac{281}{192}\, x^2 
+ \biggl( \frac{1}{4x} + \frac{1}{2} 
- \frac{15}{16}\, x \biggr) \sqrt{f} 
+\frac{3}{16}\, x (6-5x) \ln \frac{1+\sqrt{f}}{2 \sqrt{f}}, 
\label{eq:Fpoly3} 
\end{equation} 
the closed-form expression we were seeking. It is a simple matter to
verify that the expansion of the right-hand side of
Eq.~(\ref{eq:Fpoly3}) in powers of $x$ reproduces
Eq.~(\ref{eq:Fpoly2}). 

Combining Eqs.~(\ref{eq:F_large_r}), (\ref{eq:Fln2}), and
(\ref{eq:Fpoly3}), we finally obtain the closed-form expression 
\begin{subequations} 
\label{eq:F_final} 
\begin{align} 
F &= \frac{q^2 R^2}{2 r^5}\, \frac{\Xi}{f^{3/2}}, \\  
\Xi &= - \frac{1}{4x} + \frac{5}{8} + \frac{139}{96}\, x 
- \frac{281}{192}\, x^2  
+ \biggl( \frac{1}{4x} + \frac{1}{2} - \frac{15}{16}\, x \biggr) \sqrt{f} 
+\frac{3}{16}\, x (6-5x) 
  \ln \frac{\tilde{s} x (1+\sqrt{f})}{8 \sqrt{f}} 
\end{align} 
\end{subequations} 
for the electromagnetic self-force acting on a particle of charge
$q$ at a fixed radial position $r$ in the five-dimensional
Schwarzschild-Tangherlini spacetime. We have $R$ denoting the
event-horizon radius, $x = (R/r)^2$, $f = 1-x$, and 
$\tilde{s} := s/R$ is a dimensionless version of the averaging radius
introduced in the Hadamard regularization prescription.   

It is a remarkable fact that the self-force can be expressed in a
closed form obtained by summing its large-$r$ expansion. It should 
be acknowledged that these manipulations do not amount to a proof that
the self-force is indeed given by Eq.~(\ref{eq:F_final}), because the
patterns identified in the large-$r$ expansion could happen to break
at any order beyond those explicitly computed. (We have checked that
the patterns hold up to at least order $x^{34}$.) We consider this
eventuality unlikely, however, and offer the perfect agreement between
Eq.~(\ref{eq:F_final}) and the numerical data of
Sec.~\ref{subsec:numer} as evidence that we have indeed found an exact  
expression for the self-force.   

For $r/R \gg 1$, $\Xi = 1 + O(x)$, and the self-force behaves as 
$F \sim q^2 R^2/(2 r^5)$. For $r/R \to 1$, we have instead 
\begin{equation} 
\Xi = \frac{3}{16} \ln \frac{\tilde{s}}{8\sqrt{1-x}} 
+ \frac{23}{64} + O(1-x).  
\end{equation} 
This estimate can be used to obtain the bound of
Eq.~(\ref{eq:F_horizonbound}) for the self-force
near the horizon. 

\section{Scalar self-force} 
\label{sec:scalar} 

The self-force acting on a scalar charge $q$ at rest in the
five-dimensional Schwarzschild-Tangherlini spacetime can be calculated
with the same methods used to obtain the electromagnetic
self-force. Because the steps are very similar we provide a very
sparse description of this calculation. We recycle the notation
introduced in the preceding sections.  

A scalar potential $\Phi$ sourced by a scalar charge density $\mu$ in
an $(n+2)$-dimensional spacetime satisfies the wave equation 
\begin{equation} 
\Box \Phi = -\Omega_n \mu, 
\end{equation} 
where $\Box := g^{\alpha\beta} \nabla_\alpha \nabla_\beta$ is the
covariant wave operator. For a point particle with scalar charge $q$
moving on a world line $z^\alpha(\tau)$, 
\begin{equation} 
\mu(x) = q \int \delta\bigl(x,z(\tau)\bigr)\, d\tau.    
\end{equation} 
Formally, the scalar self-force acting on this particle is given by 
\begin{equation} 
F^\alpha = q \bigl( g^{\alpha\beta} + u^\alpha u^\beta \bigr) 
\nabla_\beta \Phi. 
\end{equation} 
As in the case of the electromagnetic self-force, the scalar field
$\nabla_\beta \Phi$ is infinite on the world line, and the expression
requires regularization. 

In the specific case of a static particle in the
Schwarzschild-Tangherlini spacetime, we have 
\begin{equation} 
\Box \Phi = f \partial_{rr} \Phi 
+ \frac{f}{r} (n + rf'/f) \partial_r \Phi 
+ \frac{1}{r^2} D^2 \Phi 
\end{equation} 
and 
\begin{equation} 
\mu = q \sqrt{f_0}\, \frac{\delta(r-r_0)}{r_0^n} 
\delta(\bm{\Omega},\bm{\Omega}_0). 
\end{equation} 
Placing the charge on the polar axis, a decomposition in generalized
Legendre polynomials,  
\begin{equation} 
\Phi(r,\chi) = \sum_\ell \psi_\ell(r) {\cal P}_\ell(\cos\chi), 
\end{equation} 
produces the sequence of differential equations 
\begin{equation} 
r^2 \psi''_\ell + (n + rf'/f) r \psi'_\ell 
- \frac{\ell(\ell+n-1)}{f} \psi_\ell 
= -\frac{q N(n,\ell)}{\sqrt{f_0} r_0^{n-2}}\, \delta(r-r_0) 
\end{equation} 
for the expansion coefficients $\psi_\ell(r)$. The linearly
independent solutions to the homogeneous equation are 
$\psi^{\rm in}_\ell = P_\nu(\xi)$ and 
$\psi^{\rm out}_\ell = Q_\nu(\xi)$, where $\nu = \ell/(n-1)$ and 
$\xi = 2(r/R)^{n-1} - 1$. The solution to the inhomogeneous equation
is then 
\begin{equation} 
\psi_\ell^< = \frac{2N(n,\ell)}{n-1} \frac{q \sqrt{f_0}}{R^{n-1}} 
\psi^{\rm out}_\ell(r_0) \psi^{\rm in}_\ell(r), \qquad 
\psi_\ell^> = \frac{2N(n,\ell)}{n-1} \frac{q \sqrt{f_0}}{R^{n-1}} 
\psi^{\rm in}_\ell(r_0) \psi^{\rm out}_\ell(r). 
\end{equation} 
These expressions apply even when $\ell = 0$, in which case 
$N(n,\ell) = 1$, $\psi^{\rm in}_0 = 1$, and 
$\psi^{\rm out}_0 = -\frac{1}{2} \ln f$.  

The formal self-force acting on a particle at $r = r_0$, $\chi = 0$
has $F^r = q f_0 \partial_r \Phi$ as its only nonvanishing
component. Converting to the invariant $F = f_0^{-1/2} F^r$ and
substituting the mode decomposition, we obtain 
\begin{equation} 
F = q \sqrt{f_0} \sum_\ell \psi'_\ell(r_0). 
\end{equation} 
As in the electromagnetic case, this mode sum diverges and requires
regularization. 

Our regularization prescription is based on the Hadamard Green's
function and an averaging over a small surface of constant proper
distance around the particle. A static potential in a static
spacetime with the metric of Eq.~(\ref{eq:metric_static}) satisfies  
\begin{equation} 
\nabla^2 \Phi + A^a \partial_a \Phi = -\Omega_n \mu, 
\end{equation} 
which is the same as Eq.~(\ref{eq:Phi_eq}) except for the shift $A^a
\to -A^a$ and the minus sign on the right-hand side. The associated
Green's function satisfies 
\begin{equation} 
\nabla^2 G(\bm{x},\bm{x'}) 
+ A^a \partial_a G(\bm{x},\bm{x'}) 
= -\Omega_n \delta(\bm{x},\bm{x'}), 
\end{equation} 
which is the same as Eq.~(\ref{eq:G_eq}) except for the shift $A^a 
\to -A^a$. For the case of a point charge at a fixed position
$\bm{x}_0$, the scalar potential is given by 
\begin{equation} 
\Phi(\bm{x}) = q G(\bm{x},\bm{x}_0); 
\end{equation} 
this is the same as Eq.~(\ref{eq:Phi_G2}) except for the absence of a
factor $-N(\bm{x}_0)$ on the right-hand side. 

The regularized mode sum for the self-force shall be expressed as 
\begin{equation} 
F = q \sqrt{f_0} \biggl\{ \sum_\ell \bigl[\psi'_\ell(r_0)
- \psi^{{\sf S}\prime}_\ell(r_0) \bigr]
+ \langle \partial_r \Phi^{\sf S} \rangle_{\rm ren} \biggr\}, 
\end{equation} 
and the singular potential $\Phi^{\sf S}$ shall be identified with 
$q G_{\rm H}(\bm{x},\bm{x}_0)$, where $G_{\rm H}$ is the Hadamard
Green's function. The construction of the Green's function proceeds as
in Sec.~\ref{sec:Hadamard}. In fact, there is no need to repeat any of
this work, because the scalar Green's function can be obtained
directly from the electromagnetic Green's function by implementing the
shift $A^a \to -A^a$. This observation, together with the modified
relation between $\Phi^{\sf S}$ and $G_{\rm H}$, implies that the
regularization parameters $A$, $B$, $C$, $D$, and $E$ can be obtained 
directly from Eqs.~(\ref{eq:regparam2}) and (\ref{eq:regparam4}) by
omitting the overall factor of $e^{\phi_0}$ and changing the sign in
front of $\phi_0$ and its derivatives. The relations of
Eq.~(\ref{eq:regparam1}) require no change, and the regularized mode
sum can be expressed as   
\begin{equation} 
F = q^2 \sqrt{f_0} \sum_\ell F_\ell, 
\end{equation} 
with $F_\ell$ still given by Eq.~(\ref{eq:SF_reg6}). For the specific
case of the five-dimensional Schwarzschild-Tangherlini spacetime, the
regularization parameters for the scalar self-force are given by  
\begin{subequations} 
\begin{align} 
A &= -\frac{1}{2r_0^3 \sqrt{f_0}}\,\mbox{sign}(\Delta), \\ 
B &= -\frac{2 - R^2/r_0^2}{4 r_0^3 f_0}, \\ 
C &= -\frac{R^4}{32 r_0^7 f_0}, \\ 
D &= \frac{3(2-R^2/r_0^2)R^4}{32 r_0^7 f_0^2}, \\ 
E &= \frac{5(4-R^2/r_0^2)R^4}{192 r_0^7 f_0^2}. 
\end{align} 
\end{subequations} 

We evaluate the scalar self-force as a large-$r$ expansion that will
next be summed to a closed-form expression. The large-$r$ behavior of
$\psi^{\rm in}_\ell$, obtained by combining Eq.~(8.820.4) of
Ref.~\cite{gradshteyn-ryzhik:80} with Eq.~(15.3.11) of
Ref.~\cite{abramowitz-stegun:72}, can be extracted from     
\begin{align} 
\psi^{\rm in}_\ell &= 
\frac{\Gamma(\ell+1)}{\bigl[\Gamma(\ell/2+1)\bigr]^2} (r/R)^\ell 
\biggl\{ \sum_{p=0}^\ell 
\frac{ \bigl[ (-\ell/2)_p \bigr]^2 }{p! (-\ell)_p} (R/r)^{2p} 
+ \frac{(-1)^\ell \bigl[ \Gamma(\ell/2+1) \bigr]^2} 
{\Gamma(\ell+1) \bigl[ \Gamma(-\ell/2) \bigr]^2} (R/r)^{2\ell+2}  
\nonumber \\ & \quad \mbox{} \times 
\sum_{p=0}^\infty 
\frac{ \bigl[ (\ell/2+1)_p \bigr]^2 }{p!(p+\ell+1)!} (R/r)^{2p} 
\bigl[ -2\ln(r/R) - \psi(p+1) - \psi(p+\ell+2)
+ 2 \psi(\ell/2+p+1) \bigr] \biggr\}.  
\end{align} 
The large-$r$ behavior of $\psi^{\rm out}_\ell$, obtained by
combining Eq.~(8.820.2) of Ref.~\cite{gradshteyn-ryzhik:80} with
Eq.~(15.3.20) of Ref.~\cite{abramowitz-stegun:72}, is determined by  
\begin{equation} 
\psi^{\rm out}_\ell = 
-\frac{\sqrt{\pi}\, \Gamma(\ell/2+1)}{2^{\ell+2} \Gamma(\ell/2+3/2)} 
(R/r)^{\ell+2} F(\ell/2+1, \ell/2+1;\ell+2;R^2/r^2). 
\end{equation} 
Proceeding as in the electromagnetic case, we obtain the scalar
self-force expressed as an expansion in powers of $R/r$ truncated to a
selected order $O$. For $O = 36$ we obtain 
\begin{equation} 
F = \frac{23}{96} \frac{q^2 R^4}{r^7} \biggl( F_{\rm poly} 
- \frac{18}{23} F_{\ln} \ln \frac{4 r^2}{s R} \biggr) 
\end{equation} 
with 
\begin{align} 
f^{3/2} F_{\rm poly} &= 
1 - {\frac {11}{23}}\,x + {\frac {45}{736}}\,{x}^{2}
+ {\frac {3}{64}}\,{x}^{3} + {\frac {447}{11776}}\,{x}^{4}
+ {\frac {3753}{117760}}\,{x}^{5}
+ {\frac {25863}{942080}}\,{x}^{6}
+ {\frac {63585}{2637824}}\,{x}^{7}
+ {\frac {906615}{42205184}}\,{x}^{8}
\nonumber \\ & \quad \mbox{}
+ {\frac {233575}{12058624}}\,{x}^{9}
+ {\frac {425309}{24117248}}\,{x}^{10}
+ {\frac {8587071}{530579456}}\,{x}^{11}
+ {\frac {63471735}{4244635648}}\,{x}^{12}
+ {\frac {139414947}{10032775168}}\,{x}^{13}
\nonumber \\ & \quad \mbox{}
+ {\frac {14582768229}{1123670818816}}\,{x}^{14}
+ {\frac {10522019289}{864362168320}}\,{x}^{15}
+ {\frac {45291623793}{3951369912320}}\,{x}^{16}
+ O(x^{18})
\end{align} 
and 
\begin{equation} 
f^{3/2} F_{\ln} = 1 - \frac{1}{2}\, x + O(x^{17}), 
\end{equation} 
where $x := (R/r)^2$. 

\begin{figure} 
\includegraphics[width=0.8\linewidth]{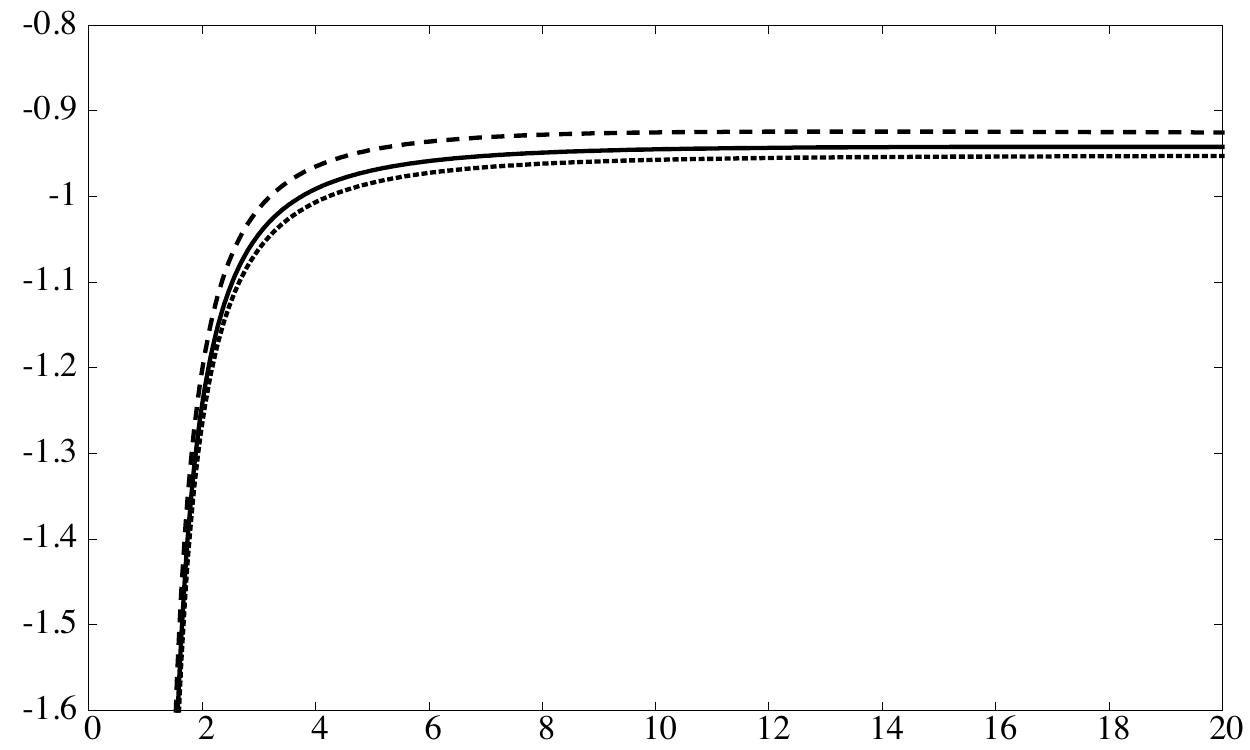}
\caption{Scalar self-force acting on a particle of charge $q$
  at position $r$ in the five-dimensional Schwarzschild-Tangherlini
  spacetime. The self-force is divided by the absolute value of 
  the asymptotic expression of Eq.~(\ref{eq:scalar_sf_asymp}), and it
  is plotted as a function of $r/R$ for $\tilde{s} = 10^{-4}$
  (long-dashed curve), $\tilde{s} = 10^{-6}$ (solid curve), and
  $\tilde{s} = 10^{-8}$ (short-dashed curve).}    
\label{fig:scalarsf} 
\end{figure} 
 
The expansion of $g_0 := f^{3/2} F_{\rm poly}$ in powers of $x$ can be
summed to a closed-form expression. The required steps are very
similar to those followed in the electromagnetic case. Letting
$s_0(p)$ be the sequence of coefficients associated with $g_0$, so
that  $g_0 = \sum_{p=1}^\infty s_0(p) x^{p-1}$, we find that the
manipulations $s_1(p) = (p-1)(p-2) s_0(p)$, $s_2(p) := s_1(p+1) 
- s_1(p)$, and $s_3(p) := s_2(p+1) - s_2(p)$ reveal a recognizable
pattern, 
\begin{equation} 
s_3(p) = \frac{135}{23} \frac{(2p)!}{2^{2p} p! (p+3)!}. 
\end{equation} 
The associated function is  
\begin{equation} 
g_3 = -\frac{9}{46 x^4} (5 x^3 - 30 x^2 + 40 x - 16) 
- \frac{72}{23 x^4} (1-x)^{5/2},  
\end{equation} 
and the original function $g_0$ is easily reconstructed. We find 
\begin{equation} 
f^{3/2} F_{\rm poly} = 
\frac{1}{92} \biggl( \frac{48}{x^2} - \frac{72}{x} + 146 - 77 x \biggr) 
- \frac{3}{23} \biggl( \frac{4}{x^2} - \frac{4}{x} + 3 \biggr) \sqrt{f} 
+ \frac{9}{23} (2-x) \ln \frac{1+\sqrt{f}}{2 \sqrt{f}}.  
\end{equation} 
With this we finally arrive at 
\begin{subequations}
\label{eq:scalar_sf} 
\begin{align}  
F &= \frac{q^2 R^4}{384 r^7} \frac{\Xi}{f^{3/2}}, \\ 
\Xi &= \frac{48}{x^2} - \frac{72}{x} + 146 - 77x 
- 12\biggl( \frac{4}{x^2} - \frac{4}{x} + 3 \biggr) \sqrt{f} 
+ 36 (2-x) \ln \frac{\tilde{s} x(1+\sqrt{f})}{8 \sqrt{f}}, 
\end{align} 
\end{subequations} 
a closed-form expression for the scalar self-force acting on a
particle of charge $q$ at a fixed radial position $r$ in the
five-dimensional Schwarzschild-Tangherlini spacetime. We recall that
$f = 1 - x$, and $\tilde{s} := s/R$ is a dimensionless version of the
averaging radius introduced in the Hadamard regularization
prescription.   
 
For $r/R \gg 1$, $\Xi = 72\ln(\tilde{s}x/4) + O(1)$, and the self-force
behaves as $F \sim F_{\rm asymp}$ with 
\begin{equation} 
F_{\rm asymp} = -\frac{3}{8} \frac{q^2 R^4}{r^7} 
\ln \frac{2r}{\sqrt{\tilde{s}} R} 
\label{eq:scalar_sf_asymp}
\end{equation} 
The self-force is {\it attractive} at large $r$, and it scales as
$R^4 \ln r /r^7$, which can be contrasted with the $R^2/r^5$ scaling
of the electromagnetic self-force. As seen in Fig.~\ref{fig:scalarsf},
the self-force stays attractive as $r$ decreases toward $R$. Near the
horizon, taking $r/R > 1 + \tilde{s}$, we find that the self-force is
bounded by   
\begin{equation} 
F > -\frac{3}{128\sqrt{2}} \frac{q^2}{R^3} 
\frac{1}{\tilde{s}^{3/2}} \ln \frac{128}{\tilde{s}},  
\end{equation} 
the same expression (\ref{eq:F_horizonbound}) that was found in the
case of the electromagnetic self-force.   

\begin{acknowledgments} 
We are grateful for discussions with Andrei Zelnikov, Valeri Frolov,
and Peter Zimmerman. This work was supported by the Natural Sciences
and Engineering Research Council of Canada.   
\end{acknowledgments}    

\appendix 

\section{Spherical harmonics in higher dimensions} 
\label{sec:Appendix} 

We collect some useful results from the literature 
\cite{hochstadt:86, frye-efthimiou:12} pertaining to the theory of
spherical harmonics in higher dimensions. The functions are defined on
a unit $n$-sphere, with $n$ denoting the number of angular directions
on the sphere. The spherical polar coordinates on the sphere are
collectively denoted $\theta^A$, with the index $A$ running from 1 to
$n$. Their relation to Cartesian coordinates $x^a$ in a Euclidean
$(n+1)$-dimensional space is given by $x^a = r \Omega^a(\theta^A)$,
where $r$ is the distance to the origin and $\bm{\Omega}$ a unit
vector in the direction of $\bm{x}$. The metric on the unit $n$-sphere
is denoted $\Omega_{AB}$, and its inverse is $\Omega^{AB}$. The area
element on the sphere is $d\Omega_n := \sqrt{\Omega}\, d^n\theta$, with   
$\Omega := \mbox{det}[\Omega_{AB}]$, and the integrated area is 
\begin{equation} 
\Omega_n := \int d\Omega_n 
= \frac{2\pi^{(n+1)/2}}{\Gamma(\frac{n+1}{2})}. 
\label{eq:volume} 
\end{equation} 
The covariant derivative operator compatible with the metric is
denoted $D_A$.      

\subsection{Definition and properties} 

The scalar harmonics $Y^{\ell,j}$ are defined in such a way that the 
function $a := r^\ell Y^{\ell,j}(\theta^A)$ satisfies Laplace's
equation $\nabla^2 a = 0$ in the $(n+1)$-dimensional space. 
The transformation to the spherical coordinates allows us to write 
\begin{equation} 
\nabla^2 a = 
\partial_{rr} a  + \frac{n}{r} \partial_r a 
+ \frac{1}{r^2} D^2 a,   
\end{equation} 
with $\partial_r$ indicating partial differentiation with respect to
$r$, and $D^2 := \Omega^{AB} D_A D_B$ denoting the Laplacian operator
on the unit $n$-sphere. Making the substitution produces the
eigenvalue equation  
\begin{equation} 
\bigl[ D^2 + \ell(\ell+n-1) \bigr] Y^{\ell,j} = 0 
\label{eq:eigenY} 
\end{equation} 
for the spherical harmonics. Here $\ell = 0, 1, 2, \cdots$, and the
index $j$ runs over a number $N(n,\ell)$ of linearly independent
functions, with  
\begin{equation} 
N(n,\ell) = \frac{(2\ell+n-1)(\ell+n-2)!}{(n-1)! \ell!}.
\label{eq:degenY} 
\end{equation} 
For $n=2$ we have the familiar $N(2,\ell) = 2\ell+1$, and for $n=3$ we
have $N(3,\ell) = (\ell+1)^2$. 

Spherical harmonics of a given degree $\ell$ can be orthonormalized by 
implementing the Gram-Schmidt procedure, while spherical harmonics
of different degrees are necessarily orthogonal. The orthonormality
relations are  
\begin{equation} 
\int \bar{Y}^{\ell,j}(\bm{\Omega}) Y^{\ell',j'}(\bm{\Omega})\,
d\Omega_n = \delta_{\ell \ell'} \delta_{j j'}, 
\label{eq:orthoY} 
\end{equation} 
with an overbar indicating complex conjugation. Here and below we use
the direction $\bm{\Omega}$ as a convenient encoding of the angles
$\theta^A$.  

Any function $b$ of the angular coordinates $\theta^A$ can be
decomposed in spherical harmonics, according to 
\begin{equation} 
b(\bm{\Omega}) = \sum_{\ell=0}^\infty \sum_{j=0}^{N-1}  
b_{\ell, j} Y^{\ell, j}(\bm{\Omega}), 
\end{equation} 
with coefficients given by 
\begin{equation} 
b_{\ell, j} = \int f(\bm{\Omega}) \bar{Y}^{\ell,j}(\bm{\Omega})\, 
d\Omega_n.  
\end{equation} 
These relations imply the completeness relation  
\begin{equation} 
\sum_{\ell j} \bar{Y}^{\ell,j}(\bm{\Omega'}) Y^{\ell, j}(\bm{\Omega})  
= \delta(\bm{\Omega},\bm{\Omega'}),
\label{eq:completeY} 
\end{equation} 
in which $\delta(\bm{\Omega},\bm{\Omega'})$ is a scalarized delta
function defined by
\begin{equation} 
\int \delta(\bm{\Omega},\bm{\Omega'}) b(\bm{\Omega'})\, d\Omega'_n 
= b(\bm{\Omega}). 
\label{eq:delta_angular} 
\end{equation} 

\subsection{Axisymmetric mode}  

Let $\chi := \theta^n$ be the angle from the polar axis, and let the
metric on the unit $n$-sphere be expressed as 
\begin{equation} 
d\Omega_n^2 = d\chi^2 + \sin^2\chi\,d\Omega^2_{n-1}, 
\end{equation} 
where $d\Omega^2_{n-1}$ is the metric on a unit $(n-1)$-sphere. We
denote by $Y^{\ell,0}$ the spherical harmonic of degree $\ell$ that
depends on $\chi$ only, and we call this function the axisymmetric
mode. Its eigenvalue equation takes the form of the ordinary
differential equation 
\begin{equation} 
\frac{d^2 Y^{\ell,0}}{d\chi^2} 
+ (n-1) \frac{\cos\chi}{\sin\chi} \frac{d Y^{\ell,0}}{d\chi} 
+ \ell(\ell+n-1) Y^{\ell,0} = 0, 
\end{equation} 
and the transformation $u=\cos\chi$ brings this to the form of a  
generalized Legendre equation 
\begin{equation} 
(1-u^2) {\cal P}''_{\ell} - n u {\cal P}'_{\ell} 
+ \ell(\ell+n-1) {\cal P}_\ell = 0,  
\label{eq:Legendre_deq} 
\end{equation} 
in which a prime indicates differentiation with $u$. The generalized
Legendre functions are defined by the generating function 
\begin{equation} 
g := (1-2ut+t^2)^{-\frac{1}{2}(n-1)} = 
\sum_\ell a_\ell {\cal P}_\ell(u) t^\ell, 
\label{eq:Legendre_generating} 
\end{equation} 
where 
\begin{equation} 
a_\ell := \biggl( \begin{array}{c} \ell + n - 2 \\ \ell \end{array}
\biggr) = \frac{(\ell+n-2)!}{(n-2)!\,\ell!}. 
\end{equation} 
When $n=2$, $a_\ell = 1$, and we recover the usual definition of the
Legendre polynomials. When $x=1$ we have that $g = (1-t)^{-(n-1)}$,
with a series expansion given by $\sum_\ell a_\ell t^\ell$;
this ensures that 
\begin{equation} 
{\cal P}_\ell(1) = 1. 
\label{eq:Legendre_norm} 
\end{equation} 
The axisymmetric mode $Y^{\ell,0}(\chi)$ and the generalized Legendre
function ${\cal P}_\ell(\cos\chi)$ differ by a normalization factor
that reconciles Eq.~(\ref{eq:orthoY}) with
Eq.~(\ref{eq:Legendre_norm}); the relation is 
\begin{equation} 
Y^{\ell,0}(\chi) = \sqrt{\frac{N(n,\ell)}{\Omega_n}} 
{\cal P}_\ell(\cos\chi). 
\label{eq:YvsP} 
\end{equation} 

\subsection{Value on the polar axis}   

Suppose that in Eq.~(\ref{eq:completeY}), the unit vector
$\bm{\Omega'}$ is aligned with the polar axis $\bm{e}$. The
completeness relation becomes   
\begin{equation} 
\delta(\bm{\Omega},\bm{e}) = 
\sum_{\ell j} \bar{Y}^{\ell,j}(\bm{e}) Y^{\ell, j}(\bm{\Omega}), 
\end{equation} 
and since $\delta(\bm{\Omega},\bm{e})$ is axisymmetric relative to
$\bm{e}$, we must have that only the $j=0$ term contributes to the
sum. This implies that $\bar{Y}^{\ell,j}(\bm{e}) 
= \bar{Y}^{\ell,0}(\bm{e}) \delta_{j0}$, and since 
$\bar{Y}^{\ell,0}(\bm{e}) = \sqrt{N(n,\ell)/\Omega_n} {\cal P}_\ell(1)  
= \sqrt{N(n,\ell)/\Omega_n}$, we have that 
\begin{equation} 
Y^{\ell,j}(\bm{e})
= \sqrt{ \frac{N(n,\ell)}{\Omega_n} } \delta_{j0}. 
\label{eq:special_value} 
\end{equation} 
Making the substitution in $\delta(\bm{\Omega},\bm{e})$, we arrive at 
\begin{equation} 
\delta(\bm{\Omega},\bm{e}) = \sum_\ell 
\frac{N(n,\ell)}{\Omega_n} {\cal P}_\ell(\cos\chi),  
\label{eq:addition2} 
\end{equation} 
where $\cos\chi = \bm{\Omega} \cdot \bm{e}$. 

\bibliography{../bib/master} 
\end{document}